\documentclass{aa}  
\usepackage{graphicx}
\usepackage{txfonts}
\usepackage{colortbl}

\begin{document} 

\title{Rossby numbers and stiffness values
     inferred from gravity-mode asteroseismology of rotating F- and B-type dwarfs}

\subtitle{Consequences for mixing, transport, magnetism, and convective penetration}

\author{C.\ Aerts\inst{1,2,3}
\and K. Augustson\inst{4,5}
\and S. Mathis\inst{4}
\and M. G. Pedersen\inst{6}
\and J. S. G. Mombarg\inst{1,7}
\and V. Vanlaer\inst{1}
\and J. Van Beeck\inst{1}
\and T. Van Reeth\inst{1}
}

\institute{
Institute of Astronomy, KU Leuven, Celestijnenlaan 200D, B-3001 Leuven, Belgium\\
email:\ {Conny.Aerts@kuleuven.be}
\and
Department of Astrophysics, IMAPP, Radboud University Nijmegen, PO Box 9010,
6500 GL, Nijmegen, The Netherlands
\and 
Max Planck Institute for Astronomy, Koenigstuhl 17, 69117, Heidelberg, Germany
\and
AIM, CEA, CNRS, Université Paris-Saclay, Universit\'e 
Paris Diderot, Sorbonne Paris Cit\'e, 91191 Gif-sur-Yvette, France
\and
Department of Engineering Science and Applied Mathematics, Northwestern
University, 2145 Sheridan Road, Evanston, IL 60208, USA
\and
Kavli Institute for Theoretical Physics, University of California, Santa Barbara, CA, USA
\and
IRAP, Universit\'e de Toulouse, CNRS, UPS, CNES, 14, avenue \'Edouard Belin, F-31400 Toulouse, France
}

\date{Received September 3, 2021; Accepted October 12, 2021}

% \abstract{}{}{}{}{} 
% 5 {} token are mandatory
 
  \abstract
   {Multi-dimensional (magneto-)hydrodynamical simulations of physical processes in stellar interiors depend on a multitude of uncalibrated free parameters, which
set the spatial and time scales of their computations.}
   {We aim to provide an asteroseismic calibration of the wave and convective
     Rossby numbers, and of the stiffness at the interface between the
     convective core and radiative envelope of intermediate-mass stars. We deduce 
these quantities for rotating dwarfs from the observed properties of their identified gravity and gravito-inertial modes.}
   {We rely on near-core rotation rates and asteroseismic models of 26 B- and 37
     F-type dwarf pulsators derived from 4-year {\it Kepler\/} space photometry,
     high-resolution spectroscopy and Gaia astrometry in the literature to deduce their convective and wave Rossby
     numbers. We compute the stiffness at the  convection/radiation interface 
     %of the convective
     %core and the radiative envelope 
     from the inferred maximum buoyancy frequency at the interface and the convective turnover frequency in the core. We use those 
asteroseismically inferred quantities to make predictions of  convective penetration levels, local flux levels of gravito-inertial waves triggered by the convective core, and of the cores' potential rotational and magnetic states.
}
{Our sample of 63 gravito-inertial mode pulsators 
covers near-core rotation rates from almost zero up to the critical
rate. The frequencies of their identified
modes lead to models with 
stiffness values between $10^{2.69}$ and $10^{3.60}$ for the
  B-type pulsators, while those of F-type stars cover the range from $10^{3.47}$ to
  $10^{4.52}$. The convective Rossby numbers derived from the maximum convective diffusion coefficient in the convective core, based on mixing length theory and a value of the mixing length coefficient relevant for these pulsators, 
  vary between $10^{-2.3}$ and $10^{-0.8}$ for B-type stars and $10^{-3}$ and $10^{-1.5}$ for F-type stars. The 17 B-type dwarfs with an asteroseismic estimate of the penetration depth reveal it to be 
  in good agreement with recent theory of convective penetration that takes into account rotation. Theoretical estimates based on the observationally inferred convective Rossby numbers and stiffness values lead to local stochastically-excited gravito-inertial wave fluxes which may exceed those predicted for non-rotating cores, in agreement with observations. Finally, the convective core of rapid rotators is expected to have cylindrical differential rotation causing a magnetic field of  $20$ to $400$\,kG for B-type stars and of $0.1$ to $3$\,MG for F-type stars.
}
{Our results provide asteroseismic calibrations to guide realistic
(magneto-)hydrodynamical simultations of rotating (magnetised) core convection in stellar interiors of dwarfs and future modelling of transport and mixing processes in their interiors.}

\keywords{Asteroseismology  --
Waves --
Convection --
Stars: Rotation --
Stars: Interiors --
Stars: oscillations (including pulsations)}

\titlerunning{Rossby numbers and stiffness values from gravito-inertial modes of rotating dwarfs}
\authorrunning{Aerts et al.}

\maketitle

%
%-------------------------------------------------------------------

\section{Introduction}

Stellar modelling of dwarfs with a convective core and a radiative envelope is
currently undergoing major advances. This boost occurs thanks to the integration
of high-precision observational constraints, notably $\mu$mag space photometry,
ever more powerful multi-dimensional hydrodynamical simulations, and new
theoretical developments \citep[see][for a review on this
integration]{Aerts2019}.  One particular aspect of the progress relates to the
new development of gravito-inertial asteroseismology applied to F- and B-type
dwarfs \citep[e.g.][for a modelling scheme]{Aerts2018}. This research field
could only emerge thanks to long-term uninterrupted photometric observations
from space. Such uninterrupted space photometry delivered the required observational
constraints to exploit long-period nonradial gravity modes, in contrast to
interrupted time-series spectroscopy or photometry assembled from ground-based
observatories. 

The best available space data for gravito-inertial asteroseismology are the
4-year duration {\it Kepler\/} light curves \citep{Koch2010}. These are assembled in
quarters of 3 months duration, after which the satellite had to be rotated to avoid sunlight
hitting the detectors. 
Stitching the 3-month-long light curve chunks from all quarters results in 4-year light curves, which deliver high-precision mode amplitudes ranging from a few to thousands of $\mu$mag.
 The low frequencies of high-amplitude 
 gravito-inertial modes are also of high
 precision given that the frequency resolution of the light curves is  $\propto 1/(1460$\,d). However, the signal at low frequencies
is only trustworthy above
$\sim\!1\,\mu$Hz. Below this value, the frequency spectra of the stitched quarter 
light curves are unreliable due to the influence
of instrumental and detrending effects \citep[see][for details]{Tkachenko2013}.
We therefore only rely on gravity or gravito-inertial
modes and waves with observed frequencies 
above $1\,\mu$Hz. Such modes are prominently present in the observed
frequency spectra of F-type and B-type dwarfs \citep[][for a review]{Aerts2021}.

While the global frequency spectra of internal gravity waves tend to be smooth
functions described by power laws \citep{Rogers2013,Alvan2014}, resonant eigenmodes with
high amplitudes occur as sharp peaks in the Fourier spectrum thanks to
efficient excitation mechanisms overcoming radiative damping \citep[see e.g.][]{Alvan2015,Lecoanet2019}. Aside from tidal
excitation \citep[ignored in this work; see][]{Fuller2017}, gravity-mode excitation in single B-type dwarfs may
occur in their thin iron convection zone due to an opacity bump in the outer
radiative envelope \citep[the so-called $\kappa$ mechanism,
e.g.][]{Pamyatnykh1999,Szewczuk2017}. Gravity-modes in F-type dwarfs occur due to flux blocking
\citep{Guzik2000,Dupret2005}. However, it has recently become clear that  gravito-inertial waves (GIWs) can also be excited stochastically by the rapidly rotating convective core because of turbulent convective Reynolds stresses and turbulent convective plumes \citep{Samadi2010,Rogers2013,Neiner2020,Augustson2020}. In the subinertial regime, waves are propagative in the convective core as inertial waves that become gravito-inertial waves in the radiative envelope \citep[e.g.][]{Mathis2014,Ouazzani2020,Lee2020,Lee2021}. In the superinertial regime, they are evanescent in the convective core and are propagative only in the radiative envelope. Stochastically-excited GIWs have recently been detected in rapidly rotating Be stars \citep{Neiner2012,Neiner2020}.
Finally, nonlinear resonant mode coupling may also be an
efficient excitation and energy exchange mechanism in dwarfs \citep[e.g.,][]{Goupil1994,Buchler1997}, with promising recent observational detections for SPB stars by \citet{VanBeeck2021}.

Irrespective of how the modes are excited, gravito-inertial asteroseismology
treats the modelling of detected and identified mode frequencies in rapidly rotating dwarfs. These
mode frequencies are often of the same order as the rotation frequency of the
star and therefore tend to occur in the subinertial regime
\citep{VanReeth2015,Moravveji2016}.  As such, asteroseismic modelling must rely
on the inclusion of the Coriolis acceleration in the theoretical pulsation
computations upon which it builds. This is nowadays done in the so-called
Traditional Approximation of Rotation 
{
\citep[TAR, e.g.,][for theoretical
descriptions suitable to compute adiabatic pulsations in the Cowling
approximation]{LeeSaio1987,Townsend2003,Mathis2009} 
}
and has proven to be highly successful in
deriving high-precision values for the masses of dwarfs (in the range of 1.3 to
9\,M$_\odot$), their evolutionary stage, their internal rotation frequency near
the convective core, and their level of envelope mixing \citep[][for a summary of asteroseismic achievements]{Aerts2021}.  

Given the recent findings from this new subfield of asteroseismology, new
theoretical developments of the TAR have emerged and are yet to be exploited by
revisiting the {\it Kepler\/} data.  The latest formulations of the TAR include
the Lorentz force due to a magnetic field \citep{Prat2019,Prat2020} and the
centrifugal acceleration for either slightly deformed stars
\citep{MathisPrat2019,Henneco2021} or in the presence of strong deformation
\citep{Dhouib2021a,Dhouib2021b}.  These new TAR-based theories offer
opportunities to go beyond the derivation of internal rotation and mixing in the
exploitation of the {\it Kepler\/} data and should allow for future probing of
internal magnetic field strengths in the presence of rotational deformation
\citep{VanBeeck2020,Mathis2021,Henneco2021,Dhouib2021a,Dhouib2021b}.

This paper gathers the currently available input from gravito-inertial
asteroseismology to guide and advance multi-dimensional magnetohydrodynamical
(MHD) simulations of dwarfs with a convective core
\citep[e.g.][]{Browning2004,Brun2005,Rogers2013}. With ever increasing
computational power, progress in such simulations is large, with the aim to
offer new understandings in transport processes in the deep interior of stars
with a convective core.  In order to be of practical use, the simulations need
to be run for proper time and spatial scales revealed by modern 4-year long
observations. 

 { Gravito-inertial asteroseismology has revealed low-frequency
  large-scale prograde modes \citep[adopting the terminology by\ ][]{Unno1989},
  also termed Kelvin modes by \citet{Townsend2003}, with frequencies between a
  few and a few tens of $\mu$Hz to be dominant in the data for the majority of
  stars \citep{Aerts2021}.  Hence, proper numerical simulations should be set up
  to treat large-scale global long-duration processes} for appropriate ranges of
the rotation frequencies of observed stars. Only few such simulations were
developed since the seminal work by \citet{Rogers2013}, given the tough
requirements on the time and spatial scales
\citep{Edelmann2019,Andre2019,Horst2020}.

While modern 3D (M)HD simulations of rotating core convection for
intermediate-mass dwarfs reveal internal gravity wave spectra in good agreement
with observations \citep{Bowman2019a,Bowman2019b,Bowman2020}, both for an anelastic approach \citep{Edelmann2019} and for
compressible fluids \citep{Horst2020}, these do not yet cover the range of 
internal rotation rates inferred for gravito-inertial pulsators. Building further on
the simulations by \citet{Browning2004}, \citet{Brun2005}, and
\citet{Featherstone2009}, \citet{Augustson2016} included both a strong core dynamo-generated
magnetic field and rotation in extensive sets of 3D MHD simulations. They
include rotation rates up to 16\% of the critical rate in the presence of
strong magnetism. However, observational data of the strongest magnetic and fast
rotating gravity-mode B-type pulsator points to a clear dominance of the
Coriolis acceleration over the Lorentz force \citep{Buysschaert2018}. Moreover, as we show below,
rotation at 16\% of the critical rate is low compared to the internal rotation frequencies inferred for most F- and B-type pulsators. We currently lack observational
estimates of the internal magnetic field strenghts of dwarfs with a convective
core, while they should be detectable from gravito-inertial modes \citep{VanBeeck2020}. 

Because the interaction of rotation and convection impacts the properties of the
magnetic dynamo, convective penetration, and wave excitation and may lead to
more rotationally-constrained regimes
\citep{Augustson2019b,Augustson2019,Augustson2020}, it is worthwhile to
reconsider 3D MHD simulations for the best setup guided by asteroseismic
constraints. Our current paper has the aim to provide such observational input
for future simulations. This input is derived from asteroseismic models based on
identified { prograde} modes detected in {\it Kepler\/} space photometry of carefully
selected pulsators. Their resonant modes allow for the derivation of three
important local time scales connected with the phenomena of convection,
buoyancy, and rotation in and near the convective core of these pulsators. We
discuss the sample of pulsators used in this work in Sect.\,2 and introduce the
relevant properties of their asteroseismic models in Sect.\,3.  Section\,4 is
devoted to the derivation of Rossby numbers and interface stiffness values,
which are important input quantities for numerical simulations. We evaluate
these observables in terms of theoretical predictions for convective penetration
and GIW fluxes in Sect.\,5. We discuss the potential rotational and magnetic
states of the convective core in Sect.\,6 and come to conclusions in Sect.\,7.

%--------------------------------------------------------------------

\section{The sample and its near-core rotation rates}

Before discussing the {\it Kepler\/} sample used in our study, it is worthwhile
to recall the broad frequency regime covered by gravity-mode asteroseismology of
dwarfs from the earliest space-based studies. The anticipated opportunity of
exploiting such modes \citep[e.g.,][]{Aerts2010} was turned into practice by
\citet{Degroote2010}. These authors discovered a period spacing pattern in the
slowly-rotating B3V star HD\,50230 from its 5-month light curve assembled with
the CoRoT space telescope \citep{Auvergne2009}. The detected pattern of period
spacings of eight dipole modes of consecutive radial order for this star is
almost constant because { it is an exceptionally slow rotator for a B-type
  star. Because of this, each of its modes can be described in terms of a
  spherical harmonic with degree $l$ and azimuthal order $m$, while the 
radial order
  $n$ denotes the overtone. 
For such a slow rotator, the mode frequencies cover a limited range. The
  detected} period spacing pattern of HD\,50230 is fully in line with the
theoretical predictions { for non-rotating stars} by \citet{Tassoul1980} and
\citet{Miglio2008}. Asteroseismic modelling ignoring the Coriolis acceleration
is appropriate for this star and allowed for the derivation of its mass, age,
and near-core envelope mixing from the small periodic deviations in the detected
pattern \citep{Degroote2010}.  These results have meanwhile been confirmed from
a more detailed independent analysis by \citet{Wu2019}, who found the level of
extra mixing in the radiative envelope to be well described by a diffusion
coefficient of $D_{\rm mix}\simeq\,10^{3.8}$\,cm$^2$\,s$^{-1}$ for this
exceptionally slow rotator among B-type gravity-mode pulsators
\citep{Pedersen2021}.  The CoRoT mission was also a pioneer in the discovery of
stochastic gravito-inertial modes subjected to the strong Coriolis acceleration
in the rapidly-rotating Be star HD\,51452 \citep{Neiner2012}. In that case, the
frequencies in an inertial reference frame were spread across a large range and
led to an assessment of convective penetration for that star.

{ These two early CoRoT applications revealed that gravity and
  gravito-inertial mode asteroseismology covers rotational frequencies from
  almost zero to almost the critical rate of the pulsators.  For rapid rotators,
  the angular dependence of each mode can no longer be expressed by a single
  spherical harmonic.  Rather, the eigenfunctions under the TAR 
are described by Hough 
functions.
  \citet{LeeSaio1997} introduced a convenient labeling system to treat
  gravito-inertial modes and Rossby modes by one set of indices
  $(k,m)$, with $k=l-|m|\geq 0$ for gravito-inertial modes and $k<0$ for 
Rossby modes.  The potential of any asteroseismic inference is completely
  determined by the ability to label each observed eigenfrequency with the
  $(k,m,n)$ of the mode causing it. This is challenging when the Coriolis
  acceleration cannot be ignored, as is the case for the gravito-inertial modes.
  Achievement of proper mode identificaton and labeling from light curves that
  cover only a few months (as in the CoRoT case) is difficult for fast rotators,
  particularly for stochastic modes.  For this reason, we here limit ourselves
  to pulsators for which identification of $(k,m)$ has been achieved without any
  ambiguity.}  Such is the case for a sample of gravity and gravito-inertial
pulsators with resonant modes discovered from the 4-year {\it Kepler\/}
\citep{Koch2010} light curves.  Thanks to the long time base of 4 years, these
light curves are so far the best suited ones to reveal gravity-mode period
spacing patterns caused by resonant modes.

Once period spacing patterns are recognised, 
they offer an optimal tool to identify the $(k,m)$ values of the modes 
from asymptotic relationships valid for  high-order gravity modes
{ treated in the TAR} \citep{Bouabid2013,VanReeth2016}.{\footnote{We recall
    that the TAR assumes the restoring buoyancy force to be much stronger than
    the Coriolis acceleration in the direction of the stable entropy and
    chemical stratifications. This allows the wave equation to become separable
    while this is not the case in general
    \citep[e.g.][]{Dintrans2000,Mathis2014}. The TAR has been confronted with
    direct 2D computations of adiabatic oscillation modes
    \citep{Ouazzani2017}. This has proven its robustness when probing the
    rotation of the radiative envelope, while it should be abandoned when
    determining the rotation inside the convective core
    \citep{Ouazzani2020,Saio2021} and evaluating stochastically-excited GIWs
    \citep{Mathis2014,Neiner2020}.}} 
Many such period spacing
patterns have been detected from {\it Kepler\/} light curves for both F-type
$\gamma\,$Dor pulsators \citep[e.g.,][]{VanReeth2015,GangLi2019,GangLi2020} and
slowly pulsating B (SPB hereafter) stars
\citep[e.g.,][]{Papics2017,Szewczuk2021,Pedersen2021}. As we will be relying on
properties of forward asteroseismic models for our research, 
we restrict ourselves here to
those $\gamma\,$Dor and SPB stars that have meanwhile been followed up and
analysed in a homogeneous way from high-resolution spectroscopy and Gaia DR2
astrometry \citep{Gebruers2021} and for which forward asteroseismic modelling of
their identified modes has been done. These requirements are fulfilled
for 37 $\gamma\,$Dor stars \citep{Mombarg2019,Mombarg2021} and 26 SPB stars
\citep{Pedersen2021}. These studies revealed best forward models with low values
of extra turbulent envelope mixing for the 37 $\gamma\,$Dor stars
($D_{\rm mix} < 10$\,cm$^2$\,s$^{-1}$) while a wide variety of 
such extra mixing was derived for the 26
SPB stars (with diffusion coefficients $D_{\rm mix}$ between roughly 10 and
$10^6$\,cm$^2$\,s$^{-1}$). For these SPB stars, the inferred $D_{\rm mix}$ 
values are mildly correlated with the near-core rotation
frequency \citep{Pedersen2021}. 

%hier spin

Various methods have been developed to deduce the near-core rotation rates,
denoted here as $\Omega_{\rm rot}\equiv 2\pi\nu_{\rm rot}$, 
from the slope of detected period spacing patterns of
identified modes
\citep{VanReeth2016,Ouazzani2017,Christophe2018,Takata2020}, all of which
leading to consistent results \citep[e.g.,][for a comparison]{Ouazzani2019}.
Here, we rely on the method { originally}
designed by \citet{VanReeth2016} { and meanwhile
updated with the improved asymptotic
  expressions for the eigenvalues of Laplace's tidal equations by
  \citet{Townsend2020}. }
  It is noteworthy that
gravito-inertial modes have their dominant probing power in the transition
region between the convective core and the radiative envelope and  therefore lead to stellar properties that are most optimally and 
quite robustly determined in that transition region \citep[e.g.][for a general review on the probing power
of various types of modes]{Aerts2021}.  Envelope and/or surface rotation
frequencies are also available for some gravity-mode pulsators \citep{Kurtz2014,Saio2015,Triana2015}, revealing
low levels of differentiality of a few percent.
Because ratios between
envelope and (near-)core rotation are not yet available for many stars
\citep[][for an overview]{Aerts2019} and given that we focus on rotating core
convection and the transition layer towards the radiative envelope, we use the
rotation rates deduced for the near-core region of the stars in the rest of this paper. 

The near-core rotation frequencies $\Omega_{\rm rot}$ and spin parameters,
$2\Omega_{\rm rot}/\omega_{n,l,m}$ with $\omega_{n,l,m}$ the angular mode
frequencies in the co-rotating frame of reference, were already deduced for the
37 $\gamma\,$Dor stars considered here \citep{Aerts2017}. \citet{Pedersen2021}
took the modelling approach one step further, using the values of
$\Omega_{\rm rot}$ derived from the method by \citet{VanReeth2016} as initial
estimate to optimise its value from a 7D modelling framework following
\citet{Aerts2018}. The values obtained by \citet{Pedersen2021} are used for the
26 SPB stars in our sample, while we took their spin parameters from the
follow-up study by \citet{Pedersen2021b}.  { We note that several of these 26
  SPB stars and most of the 37 $\gamma\,$Dor stars have spin parameters 
too high
  to represent their eigenfunction by the single spherical harmonic
  $Y_1^1$. This spherical harmonic component delivers the dominant contribution
  to the Hough eigenfunction for spin parameters up to $\approx
  7$. Nevertheless, these low-frequency prograde modes have been called
  ``dipole'' prograde modes for simplicity in the recent literature.}

\begin{figure}[h!]
\begin{center} 
\rotatebox{270}{\resizebox{7cm}{!}{\includegraphics{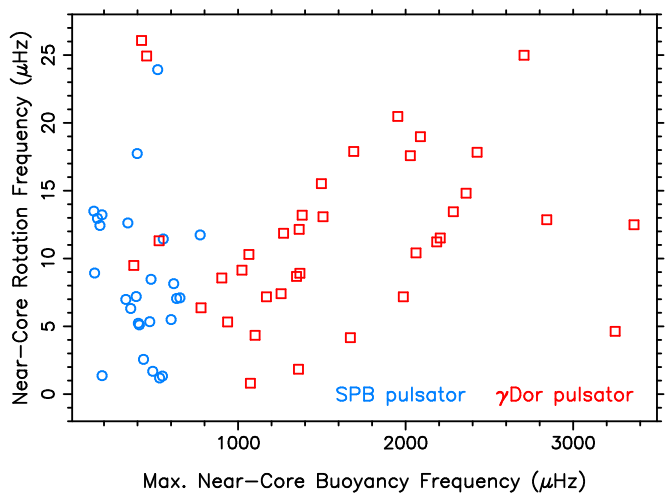}}}
\end{center}
\caption{\label{BV-Frot} Near-core rotation frequencies versus the maxima
  of the buoyancy frequency profile in the transition region region between the
  convective core and the radiative envelope (cf.\ Fig.\,\ref{Turnover}) for a
  sample of 37 $\gamma\,$Dor (red squares) and 26 SPB (blue circles) pulsators.}
\end{figure}

The cyclic near-core rotation frequencies, $\nu_{\rm rot}$, of the 63
gravity-mode pulsators in our sample are shown in Fig.\,\ref{BV-Frot}, where
they are compared with the maximum of the cyclic buoyancy frequency profile in
the transition layer between the convective core and the radiative envelope
($N_{\rm max}$).  The ratios $N_{\rm max}/2\nu_{\rm rot}$ deduced from
Fig.\,\ref{BV-Frot} range from 5.2 to 223 for the SPB pulsators and from 8.1 to
666 for the $\gamma\,$Dor stars, with median values of 28 and 68, respectively.
We will use these values in Sect.\,4 to compute the Rossby numbers but for now
point out that the 63 stars in our sample cover the entire core-hydrogen burning
phase. Moreover, the cyclic near-core rotation frequencies shown in
Fig.\,\ref{BV-Frot} range from almost zero (nonrotating case) to the critical
rotation frequency \citep[see][for detailed discussions of these coverages and
relationships between derived properties and evolutionary
status]{Mombarg2021,Pedersen2021}.  { It is noteworthy to recall that the
  critical rotation frequencies of our target stars have been derived in the
  simplistic Roche model, which assumes that the potential of the star can be
  computed as if its entire mass is concentrated at the stellar centre
  \citep[e.g.,][]{Georgy2013}.  However, the sample stars rotating close to
  their critical rate are quite deformed due to the centrifugal acceleration.  Despite
  this, all the 63 stars in our sample were modelled with the TAR in its
  original form, which neglects the centrifugal deformation. Future improvements
  can be made from the recent TAR upgrades for moderate to strong deformation 
by \citet{MathisPrat2019,Henneco2021,Dhouib2021a,Dhouib2021b}. }

\section{Properties of gravito-inertial asteroseismic models}

\begin{figure}
\begin{center} 
\rotatebox{270}{\resizebox{6.2cm}{!}{\includegraphics{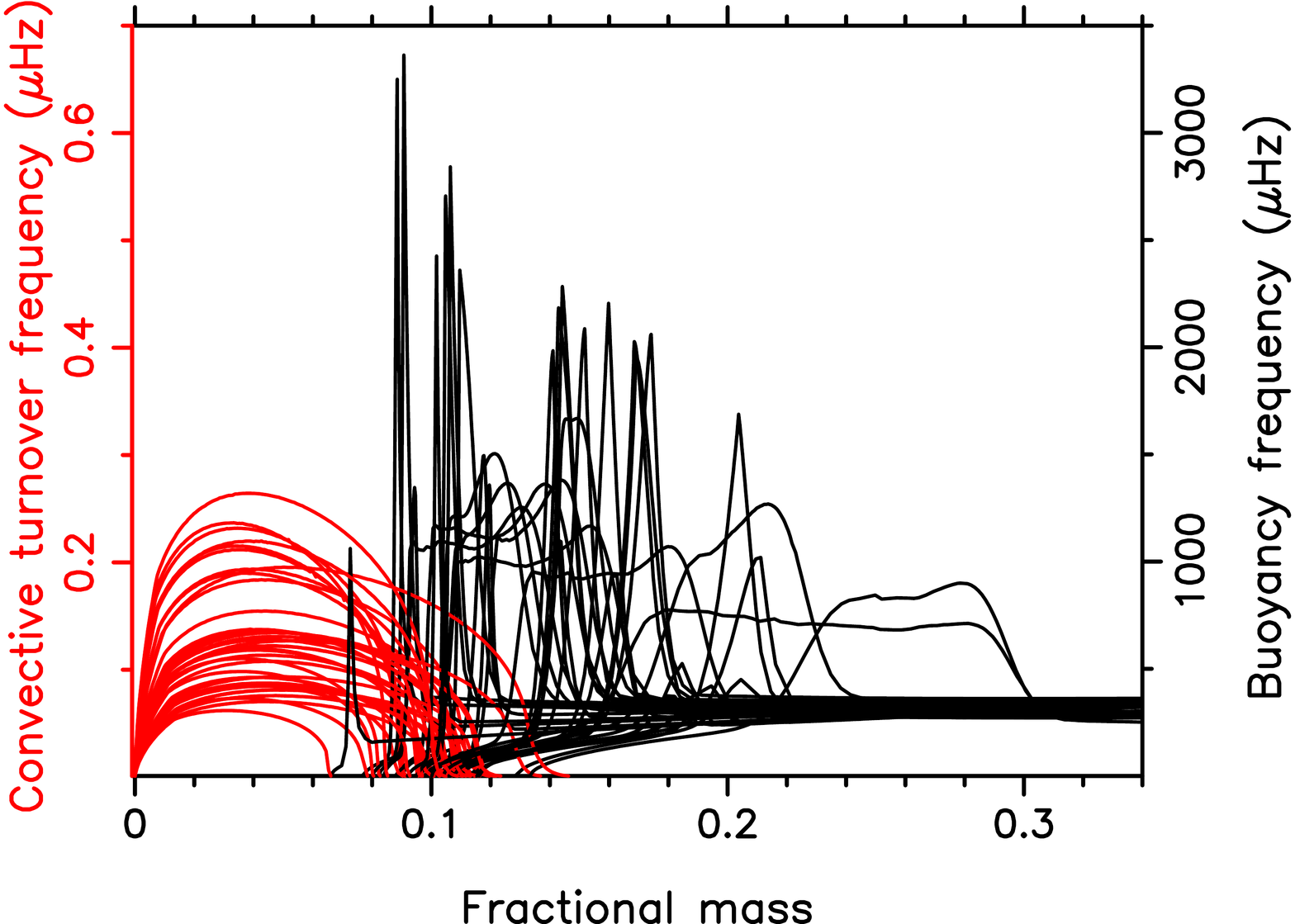}}}\\[0.2cm]
\rotatebox{270}{\resizebox{6.2cm}{!}{\includegraphics{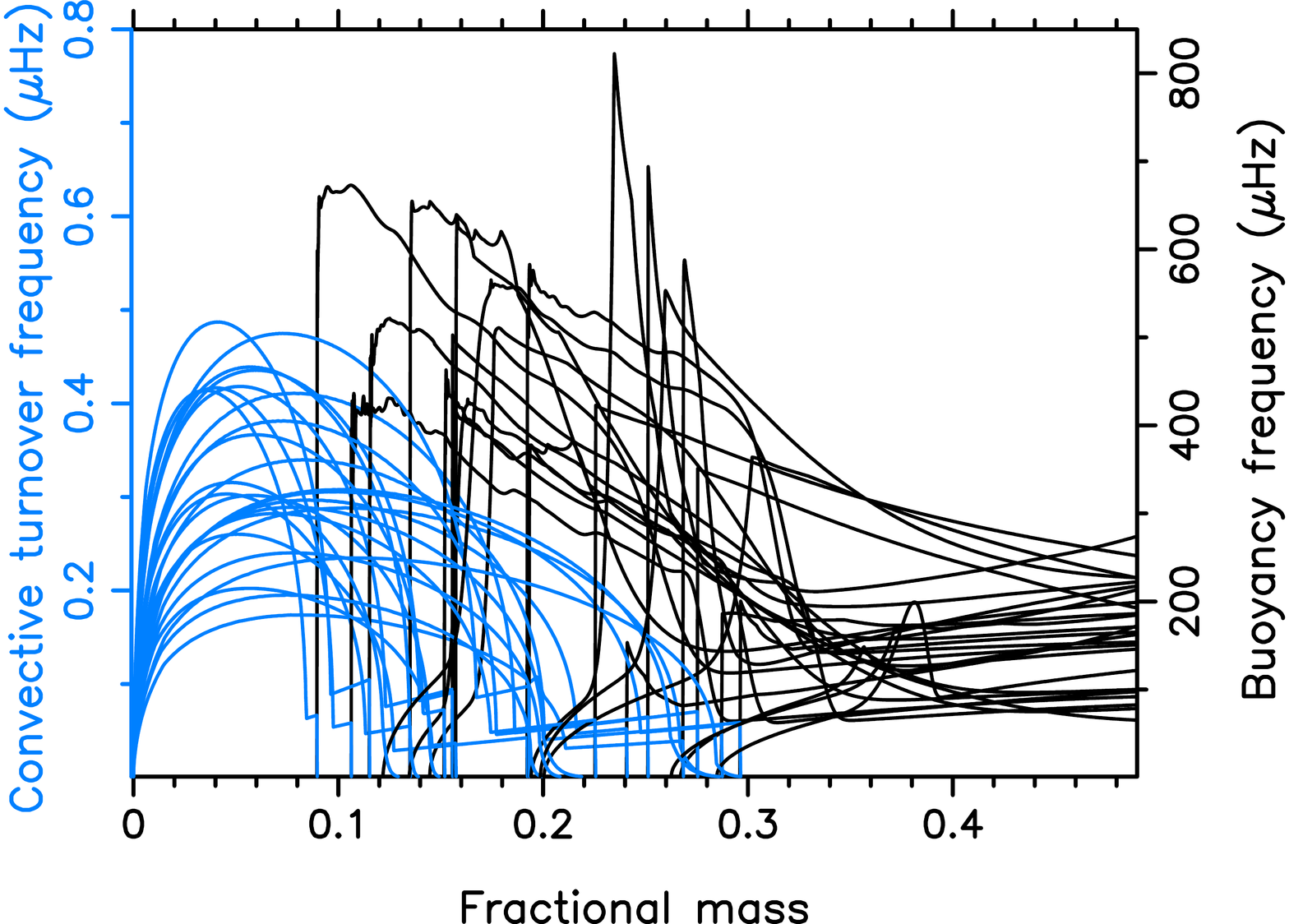}}}
\end{center}
\caption{\label{Turnover} Profiles of the cyclic 
buoyancy frequency (in black; right $y$ axis) and of the cyclic
  characteristic convective turnover 
  frequency (coloured; left $y$ axis) in the
  transition region between the core and the radiative envelope of the 
37 $\gamma\,$Dor stars (upper panel) and 26 SPB stars (lower panel) 
plotted as a function of fractional mass.}
\end{figure}

The modelling of gravity and gravito-inertial
modes and waves is already described in detail in the literature \citep[][for a review]{Aerts2021}. We stress the importance of including the Coriolis acceleration in the modelling, in order to get meaningful results for the majority of observed gravity-mode pulsators.
Because this is sometimes ignored in recent studies \citep[e.g.,][]{Wu2018,Lecoanet2019}, we recall that considering the Coriolis acceleration as a small perturbation
already breaks down for projected surface rotation velocities of $v\sin\,i\simeq\,30\,$km\,s$^{-1}$ \citep{Schmid2016}.

Based on the
subinertial nature of most of the detected gravity modes \citep{Aerts2017},
asteroseismic forward modelling frameworks relying on the TAR were developed and applied to rotating dwarfs
\citep[e.g.][]{Moravveji2016,VanReeth2016,Aerts2018,Szewczuk2018,Mombarg2019}. 
Here, we rely on the best forward models based on the TAR available for the 63
stars in our sample. These are taken from the sample studies by \citet{Mombarg2021} and 
\citet{Pedersen2021}. The $\gamma\,$Dor stars in this sample were so far only modelled with diffusive overshooting adopting the radiative temperature gradient in the overshoot zone. 
As discussed in \citet{Mombarg2019} and \citet{Mombarg2021} it was not possible to discriminate between a step or exponentially decaying overshooting prescription in the transition layer between the core and the envelope. Moreover, the values of the diffusive mixing coefficient in the envelope beyond the overshoot zone as probed by the gravity and gravito-inertial modes are very low ($D_{\rm mix}$ below 10\,cm$^2$\,s$^{-1}$) for these pulsators when ignoring radiative levitation and shear instabilities. On the other hand, discrimination between convective penetration and diffusive overshooting was achieved for the SPB stars by \citet{Pedersen2021}.  These authors inferred high values of the mixing coefficient in the radiative envelope of such pulsators, spanning five orders of magnitude for the 26 stars. 
We show the buoyancy profiles in the inner part of these 63 best asteroseismic stellar
models in Fig.\,\ref{Turnover}. It is seen from these profiles that the sample
covers a wide variety of stars in terms of evolutionary stage, where several
stars have strongly receded convective cores that have left behind broad
molecular weight gradient zones contributing to the buoyancy profile. On the other
hand, some of the $\gamma\,$Dor stars are in the mass regime and evolutionary
stage where they have growing convective cores \citep{Mombarg2019}.  We already
used the maxima of the profiles shown in Fig.\,\ref{Turnover} to construct 
Fig.\,\ref{BV-Frot}.

Further, we show in Fig.\,\ref{Turnover} the profiles of the characteristic
frequency of convection in the core deduced from the overall estimation of the diffusive mixing coefficient and the adopted mixing length parameter. The characteristic convective frequency 
shown in Fig.\,\ref{Turnover} was
computed as
\begin{equation}
\nu_{\rm char}^{\rm conv} = \frac{v_{\rm conv}}{\ell_{\rm MLT}},
\end{equation}
where $v_{\rm conv}$ is the local convective velocity and
$\ell_{\rm MLT}=\alpha_{\rm MLT}\cdot H_p$ is the local length scale over which
the fluid elements travel before dissolving, when adopting the mixing length
theory of convection \citep{BohmVitense1958} with $H_p$ the local pressure scale
height. For the $\gamma\,$Dor stars, the value of the mixing length parameter,
$\alpha_{\rm MLT}$ was taken to be 1.73 \citep{Choi2016}.  This is the solar
value for the frozen envelope input physics adopted by \citet{Mombarg2021} in
their modelling of the $\gamma\,$Dor stars in our sample.  \citet{Pedersen2021}
took a value of $\alpha_{\rm MLT}=2.0$ for the SPB stars given that their input
physics adopted for the envelope mixing is not meaningful as physical ingredient
for solar-scaled models.  This value of 2.0 is a typical value for
intermediate-mass stars deduced from the red edge of the classical instability
strip based on time-dependent convection in comparison with $\delta\,$Sct
pulsators \citep{Dupret2005}.  
{ The temperature gradient in the convective core is essentially the
  adiabatic one, hence the size of the convective core 
is insensitive to the value of $\alpha_{\rm
    MLT}$. This is also the case for the size and the structure 
of the radiative envelope, where the gravito-inertial waves propagate.
Moreover, the value of  $\alpha_{\rm MLT}$ in the convective core }cannot be
calibrated from gravity or gravito-inertial modes computed within the TAR,
because these modes do not propagate in the core. This
inability was demonstrated by both \citet{Aerts2018} and \citet{Johnston2019},
who showed that changing from $\alpha_{\rm MLT}=1.0$ to $\alpha_{\rm MLT}=2.2$
leads to stellar models with indistinguishable mode frequencies for 1-year light
curves and buoyancy travel times ($\Pi_0$), respectively. A potential way to
estimate $\alpha_{\rm MLT}$ could be to adopt a non-TAR treatment of GIWs for
which subinertial GIWs convert into propagative inertial waves in the convective
core \citep{Dintrans2000,Ouazzani2020,Saio2021}.

From Fig.\,\ref{Turnover}  we extracted the maximum value of the characteristic
frequency of convection for each of the stars. The values range from 0.17 to
0.49\,$\mu$Hz for the SPB stars and from 0.062 to 0.264\,$\mu$Hz for the $\gamma\,$Dor
stars. We stress that these characteristics frequencies are model dependent and are determined by the adoption of the mixing-length theory. These values result from the best available forward asteroseismic model of the 63 pulsators, assuming the mixing length theory of convection with the most reasonable values for $\alpha_{\rm MLT}$. This does not exclude the existence of other stellar models with a different value of $\alpha_{\rm MLT}$ that are also capable of fitting the observed and identified modes in these stars.

\section{Rossby numbers and interface stiffness values}

\begin{figure}
\begin{center} 
\rotatebox{270}{\resizebox{6.4cm}{!}{\includegraphics{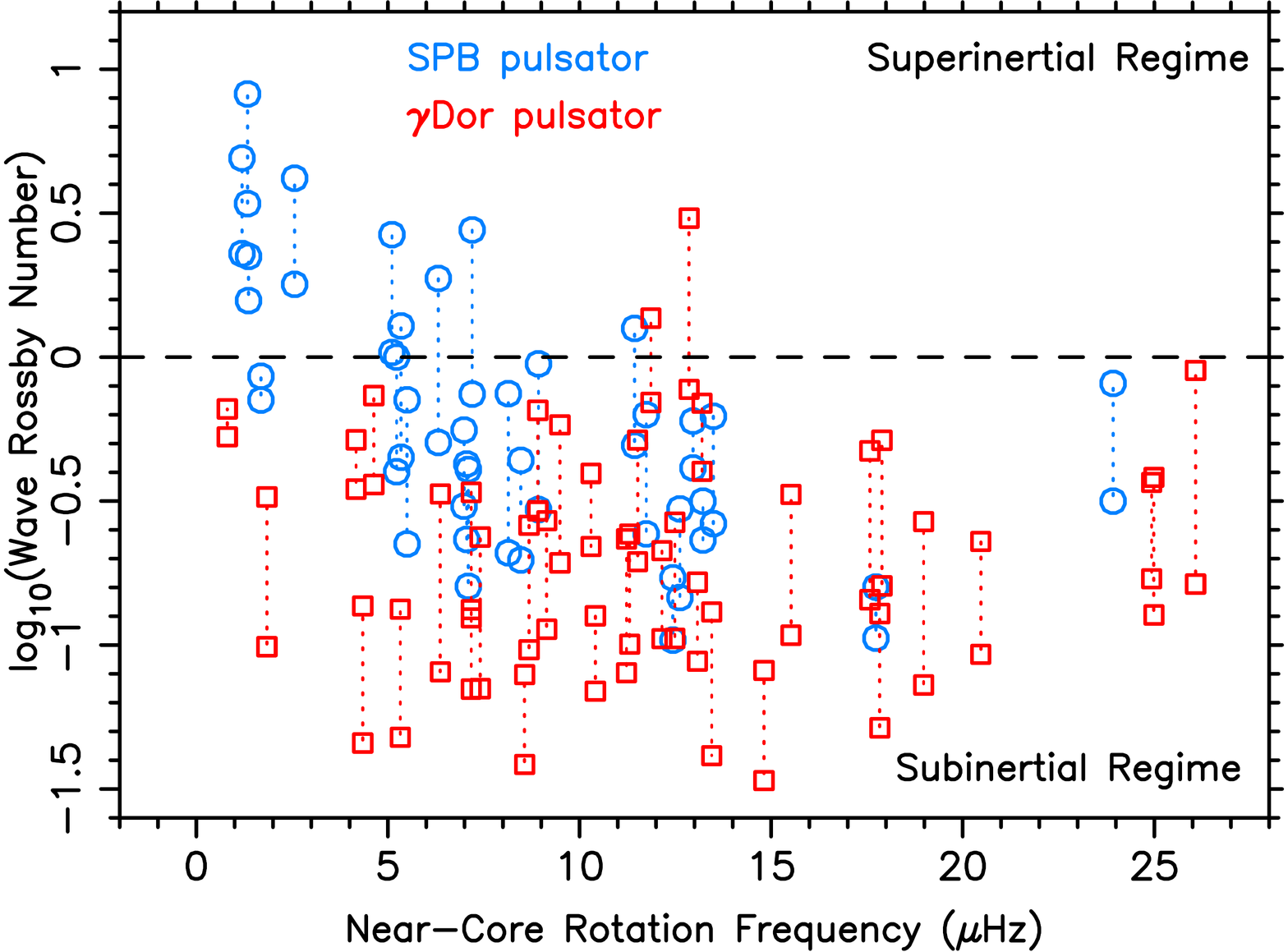}}}
\end{center}
\caption{\label{wave-rossby} Wave Rossby numbers computed as the inverse of the
  spin parameters for the sample stars. For each of the stars, the minimal and maximal
  wave Rossby numbers of the identified { low-frequency prograde} 
modes of each star are connected by a dotted vertical line. The horizontal
  dashed line distinguishes the superinertial and the subinertial regime of the waves.}
\end{figure}

Rossby numbers constitute key quantities for simulations and theoretical
predictions of wave spectra and properties \citep{Augustson2020}. Figure\,\ref{wave-rossby} shows the
wave Rossby numbers, defined as the inverse spin parameters 
\begin{equation}
{\rm Ro}_{\rm w}=\omega_{n,l,m}/2\Omega_{\rm rot}
\end{equation} 
of the sample stars. In practice, we took all the identified 
{ low-frequency prograde}
modes occurring in the period spacing patterns detected by \citet{VanReeth2015} for the $\gamma\,$Dor stars and by \citet{Pedersen2021}
for the SPB stars. For all these identified modes, we computed the frequency in the corotating frame from the star's rotation frequency $\Omega$ deduced from the forward modelling by \citet{VanReeth2016} and \citet{Pedersen2021}, respectively. This leads to a list of spin parameters per star based on each of its identified modes. These spin parameters were already computed by \citet{Aerts2017} for the $\gamma\,$Dor stars and were taken from \citet{Pedersen2021b}
for the SPB stars. 
In Fig.\,\ref{wave-rossby} we connected the lowest and highest of all the Rossby numbers per star by a dotted vertical line segment. These line segments are hence 
observables based on the minimal and
maximal spin values of the identified { low-frequency prograde} 
modes for each of the stars in our sample.  This covered range of wave Rossby numbers per star is plotted against the
near-core rotation frequency per star.  Figure\,\ref{wave-rossby}  illustrates 
that almost all modes in the
$\gamma\,$Dor stars are subinertial. The detection of such modes could be an asset to probe the possibly differential rotation in the core \citep{Ouazzani2020,Saio2021}, but this potential has yet to be put into practice.
For the SPB stars about half of the modes
are superinterial and half subinertial. This, along with lower rotational
frequencies, led to Rossby modes being detected in the $\gamma\,$Dor stars
\citep{VanReeth2016,Saio2018} but not in the SPB stars \citep{Pedersen2021}. 

\begin{figure}
\begin{center} 
\rotatebox{270}{\resizebox{6.4cm}{!}{\includegraphics{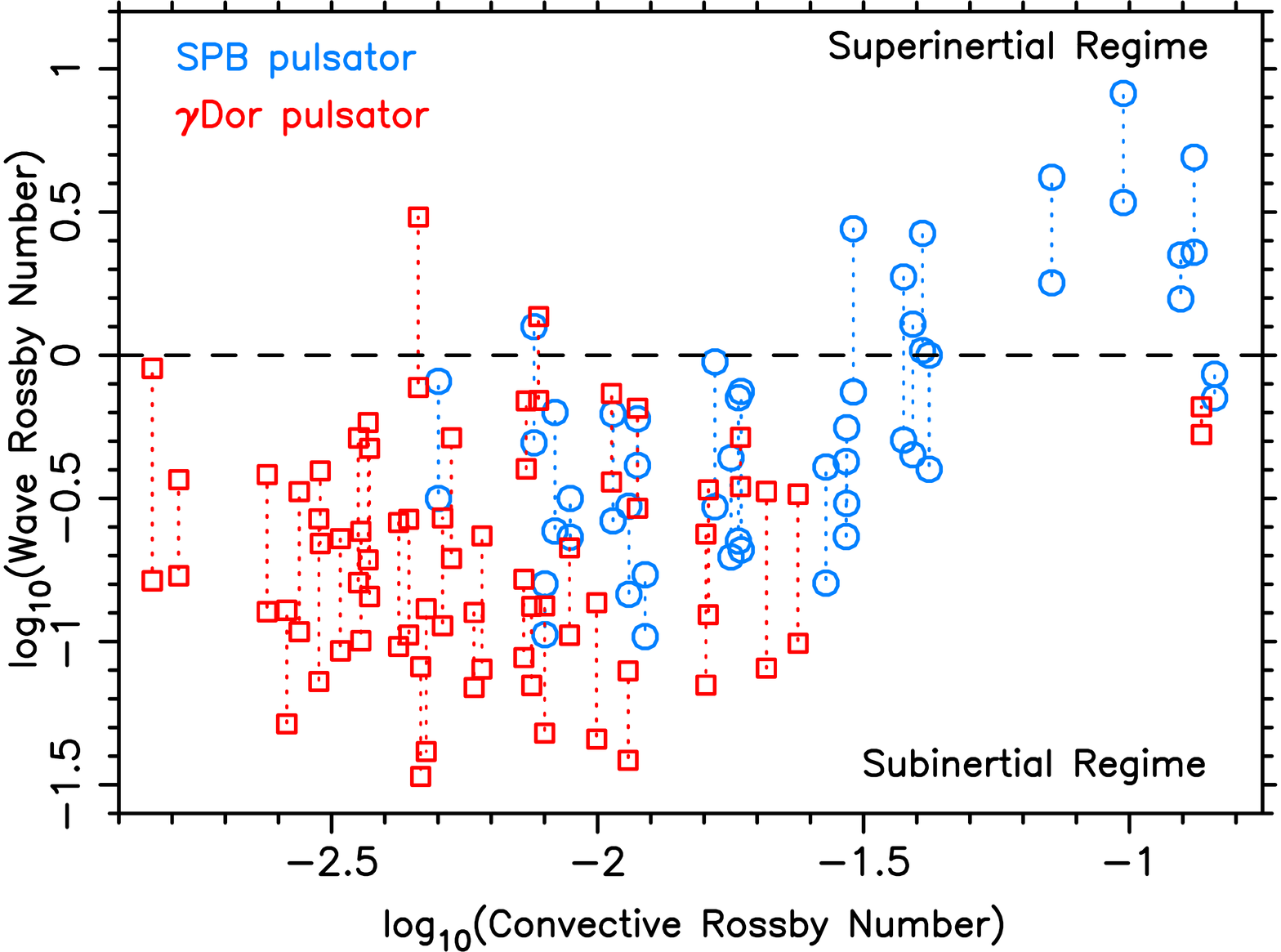}}}
\end{center}
\caption{\label{conv-rossby} Wave Rossby numbers plotted against the convective
  Rossby numbers computed as  $\nu_{\rm char}^{\rm conv}/2\nu_{\rm rot}$ deduced from the maxima inside the convective cores shown in Fig.\,\ref{Turnover}. 
The horizontal
  dashed line distinguishes the superinertial and the subinertial regime of the waves.}
\end{figure}

Aside from the wave Rossby numbers, we also computed the convective Rossby
numbers. Given that our work relies on the outcome of forward asteroseismic modelling, we approximate the convective Rossby numbers as 
\begin{equation} 
{\rm Ro}_{\rm c}=\nu_{\rm char}^{\rm conv}/2\nu_{\rm rot} 
\end{equation} and show
their values in comparison with the wave Rossby numbers in
Fig.\,\ref{conv-rossby}.  This figure reveals that the two groups of stars
occupy somewhat different regions, with the $\gamma\,$Dor stars having
lower convective Rossby numbers than the SPB stars. Following
\citet{Augustson2020}, this makes it harder for GIWs to
get excited with strong flux at the core/envelope interface for $\gamma\,$Dor stars
compared to SPB stars. This conclusion does not change if we compute the convective Rossby numbers from an average value of the characteristic convective turnover frequency in the core region rather than taking its maximum value inside the core. Such averaging roughly lowers this characteristic turnover frequency by a factor of two, implying a shift in the convective Rossby numbers (plotted as $x-$values) of about $-0.3$. In this sense, Fig.\,\ref{conv-rossby} provides an upper limit of the convective Rossby numbers under the assumption that core convection is well described by mixing length theory with the adopted mixing length scale.

\begin{figure}
\begin{center} 
\rotatebox{270}{\resizebox{6.4cm}{!}{\includegraphics{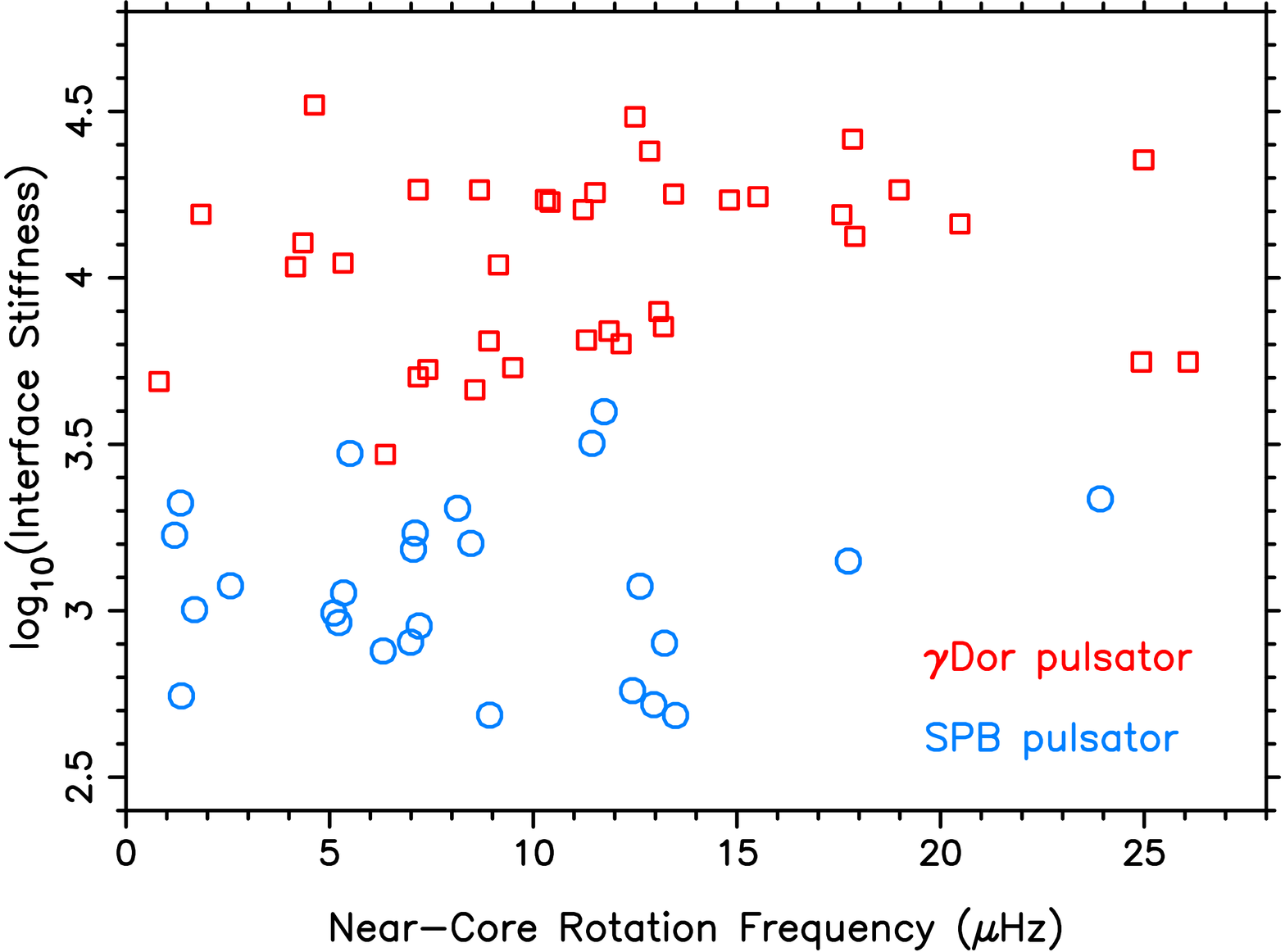}}}
\end{center}
\caption{\label{Stiffness} Ratio of the buoyancy frequency in the stable region to the convective turnover frequency in the convective core (interface stiffness) plotted as a function of
  the near-core rotation frequency for the sample of gravito-inertial pulsators.}
\end{figure}

Following \citet{Augustson2020}, we define the stiffness, $S$, at the interface
between the convective core and the radiative envelope to be the ratio of the
maximum of the buoyancy frequency outside of the convective core 
(cf.\,Fig.\,\ref{Turnover})
and the maximum characteristic convective turnover frequency in the core. Other
studies use the terminology of bulk Richardson number for this dimensionless quantity
instead of stiffness \citep[e.g.,][]{Christini2019,Scott2021}.  
We show the values of $S$ for the 63 stars in our sample in Fig.\,\ref{Stiffness}. The
$\gamma\,$Dor pulsators have higher stiffness values (roughly between 3\,000 and 35\,000) than the SPB stars (roughly between 500 and 4\,000).
These stiffness values are in line with those
considered by \citet[][their Fig.\,6]{Augustson2020} in their prediction of the
local flux of GIWs triggered by the rotating convection in
their models of intermediate-mass stars, as discussed in the next section.

\section{Theoretical predictions for convective penetration and gravito-inertial wave flux}

\subsection{Sizes of convective penetration zones}

\begin{figure}
\begin{center} 
\rotatebox{270}{\resizebox{6.5cm}{!}{\includegraphics{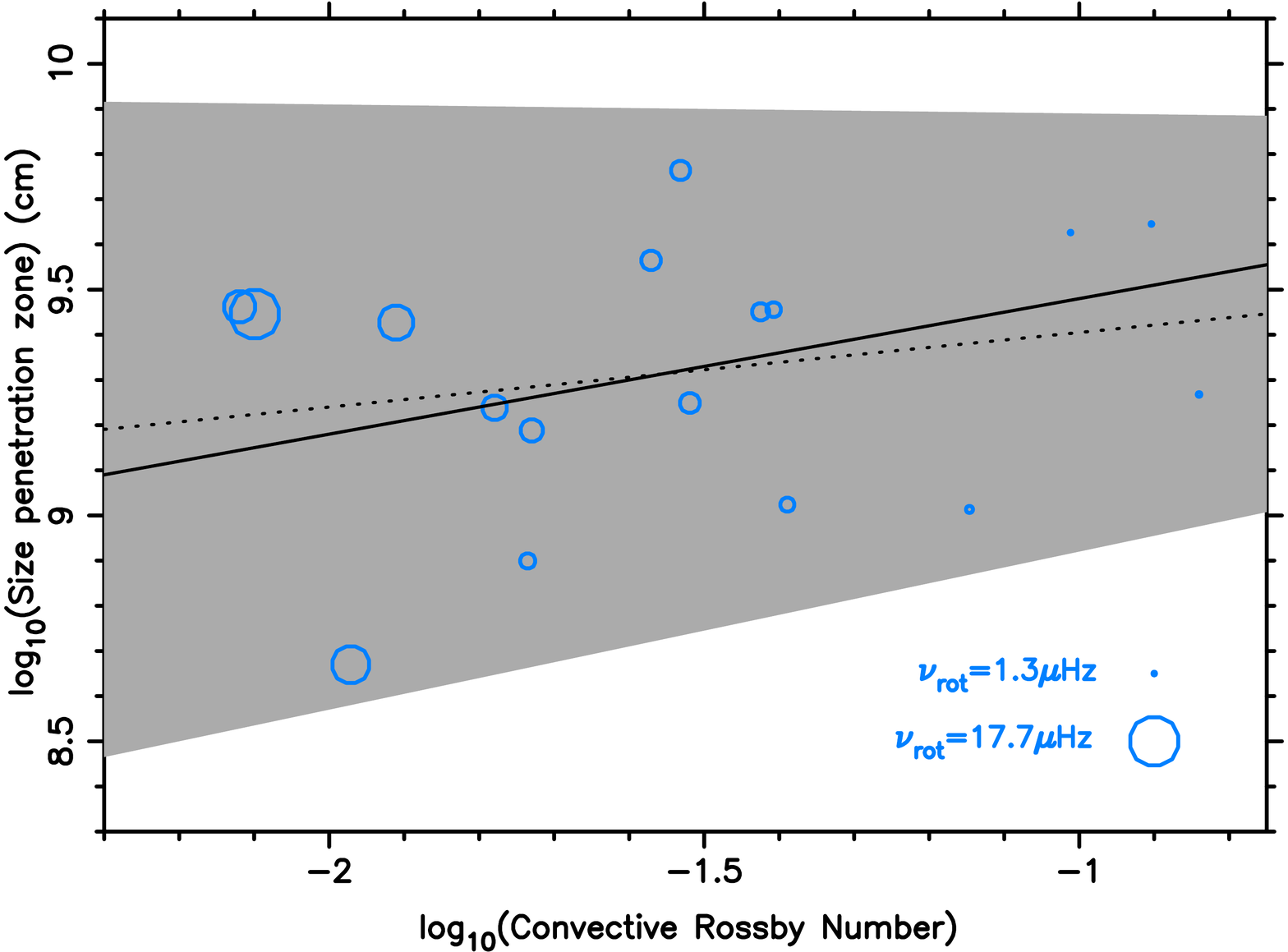}}}
\end{center}
\caption{\label{Depth} Size of the penetration zone plotted as a function of convective Rossby number for the 17 SPB pulsators in \citet{Pedersen2021} whose best forward asteroseismic model is based on convective penetration (the symbol sizes scale linearly with the near-core rotation frequency, where the legends list the lowest and highest values).
The full line represents the expected relationship according to the theory by
\citet{Augustson2019}, with a slope fixed at 3/10 and the intercept calibrated
by the 17 observed stars. The dotted line represents the best linear fit, with
its uncertainty region indicated in grey.  }
% note that I could indicate a right axis with the fraction of the solar radius;
% this is well accessible for a nonexpert audience
\end{figure}

In principle, gravity modes offer good probing power to assess the 
{ mass and the} 
thermal and chemical properties of the transition layer between the convective core and the radiative envelope \citep{Pedersen2018,Michielsen2019}. 
The
theory of convective penetration developed by \citet{Augustson2019} predicts the penetration depth from the core into the envelope to increase with the convective Rossby number as $L_P \approx \mathrm{Ro}_c^{3/10}$.  We are able to verify this result observationally by considering the gravity-mode pulsators whose best models are based on convective penetration. At present, none of the $\gamma\,$Dor stars have been modelled with convective penetration as it turned out to be difficult to distinguish between different temperature gradients and mixing prescriptions in the transition layer between the convective core and radiative envelope for these pulsators \citep{Mombarg2019,Mombarg2021}. The SPB pulsators studied by \citet{Pedersen2021} do offer a good test case as they have been modelled considering 
eight grids of equilibrium models based on different input physics. Four of these grids are based on convective penetration with the adiabatic temperature gradient in the transition zone between core and envelope, whereas the other four grids use diffusive overshooting, which adopts the radiative temperature gradient in that zone.
 This study revealed 17 of the 26 SPB stars to have their gravity 
modes best explained by models with convective penetration. 

We computed the sizes of the penetration zones for these 17 SPB stars from their best asteroseismic model and evaluate their relation to $\mathrm{Ro}_c$. 
The results are shown in Fig.\,\ref{Depth}, 
where the symbol sizes scale linearly with the near-core rotation frequency of the stars. The dotted line is the best univariate linear regression fit and the grey area indicates its uncertainty region. It can be seen that the predictive power of this linear model is limited. We also computed a bivariate linear fit by adding the near-core rotation frequency as covariate but this did not lead to a better model when adopting the Bayesian Information Criterion as model selection tool \citep{Claeskens2008}. The
full line in Fig.\,\ref{Depth} represents a linear fit relying on the prediction from the theory in  \citet{Augustson2019} with a slope fixed at 3/10 and is in agreement with the observational trend to within the uncertainties. 

 In order to limit the dimensionality of the challenging forward modelling
 problem, \citet{Pedersen2021} fixed the efficiency of the mixing accompanying
 the penetration according to the diffusion coefficient inside the convective
 core, close to its inner boundary. We thus do not yet have any estimate of
 this mixing efficiency as it has been fixed to a reasonable
 value. Figure\,\ref{Depth} hints towards stars with lower convective Rossby
 numbers and faster rotation having their fluid elements originating from within
 the core penetrate less deep into their envelope and thus experiencing less
 element transport in that transition zone. Given the current limited sample
 size, however, this cannot be but considered a hint that needs further
 investigation from a much larger sample, { taking into account the
   variability in mass, radius, and evolutionary stage across the sample}.

\subsection{Wave flux predictions}

\begin{figure*}
\begin{center} 
\includegraphics[width=\textwidth]{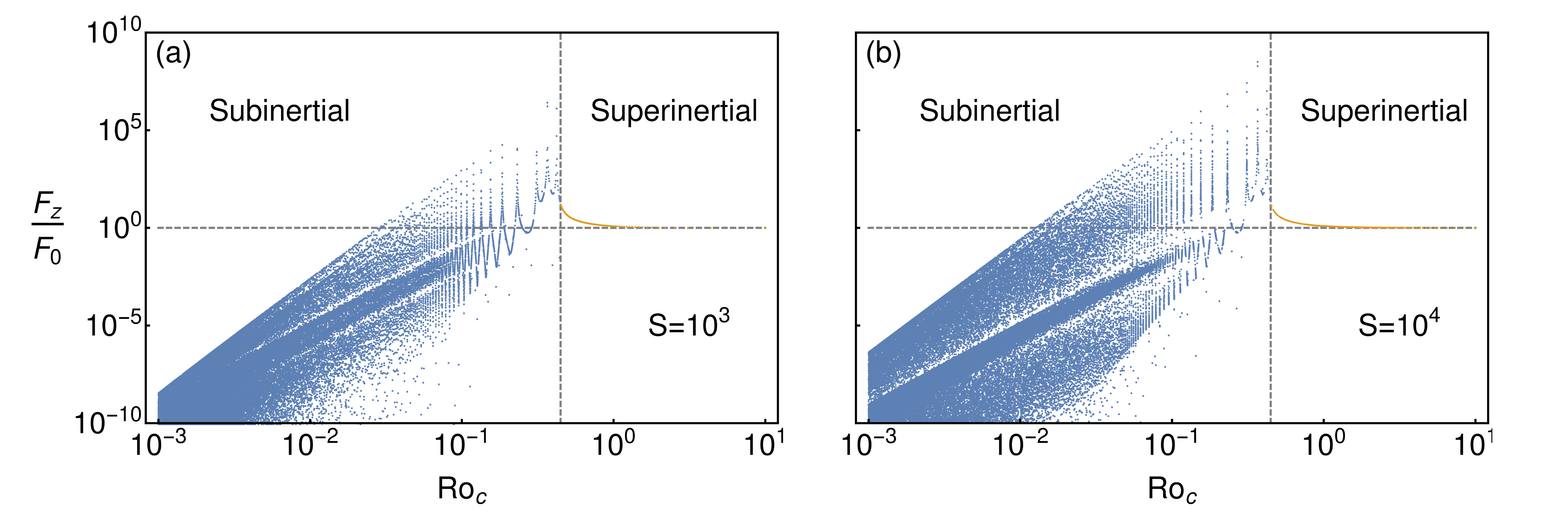}
\caption{Scaling of the gravito-inertial wave flux $F_z$, normalised by the gravity wave flux for the non-rotating case $F_0$, when excited by columnar convection at the equator for waves. The stiffness adopted in the theory by \citet{Augustson2020}, defined as the ratio of the buoyancy frequency to the convective overturning frequency ($S = N_R/N_0$ in their notation) is approximated here by $N_{\rm max}/2\nu_{\rm rot}$ from Fig.\,\ref{BV-Frot} and taken to be (a) $10^3$ for the SPB stars (left panel), (b) $10^4$ for the $\gamma\,$Dor stars (right panel) as found from Fig.\,\ref{Stiffness}. The vertical dashed line denotes the transition between subinertial and superinertial regime placed at a convective Rossby number of $\sqrt{5}/5$ above which the waves decay as indicated by the orange curve. The horizontal line denotes unity.}\label{fig:wavefluxAMA20}
\end{center}
\end{figure*}

From an observational point of view, we can only assess the GIW flux at the surface of the star. Since the variability of the 63 pulsators in our sample is dominated by coherent resonant modes, we first have to prewhiten all signal due to such significant and resolved modes before hunting for GIWs signals. 
This prewhitening has been done under the assumption that the resonant modes have infinite lifetime, constant amplitude, and constant phase, as is the common treatment for heat-driven modes. Stochastic GIWs are damped and thus have shorter lifetimes than the resonant modes of the star, implying Lorentzian-shaped frequency peaks in the amplitude spectrum, instead of the sharp delta peaks due to the resonant modes. Moreover, GIWs have time-variable amplitudes and phases.  
During the prewhitening process, any stochastic GIWs with significant amplitudes according to a chosen significance criterion will also get extracted and prewhitened, but with a wrong mathematical functionality. This leaves signal from their Lorentzian wings in the residual light curve. 

In order to assess if stochastically-excited GIWs could be present in our sample stars, we take a conservative approach. We rely on the residual light curves of the 37 $\gamma\,$Dor stars computed by \citet{VanReeth2015}. For the 26 SPB stars, we take  the residual light curve  after prewhitening from the method that gave the highest scaled fractional variance among the five methods exploited 
by \citet{VanBeeck2021}. 
We compute the amplitude spectra of these 63 residual light curves.
The signal left in those residual amplitude spectra may be caused by several phenomena, such as unresolved resonant modes, resonant modes with time-variable amplitudes (e.g.\ due to nonlinear mode couplings), leftover signal of the poorly prewhitened damped GIWs, spots subject to differential surface rotation, magnetic activity, instrumental effects, etc. For the purpose of assessing whether the leftover signal may be a manifestation of GIWs at the stellar surface, we rely on
the maximum amplitude in the residual spectrum for each of the 63 stars. 
Since there is no good way to predict the amplitudes of heat-driven resonant coherent modes, we compute the ratio of the largest-amplitude signal in the residual spectrum, $A_{\rm res}$, to the largest-amplitude coherent mode, $A_{\rm coh}$. This ratio $A_{\rm res}/A_{\rm coh}$ covers the interval $[0.09,3.64]\%$ for the 37 $\gamma\,$Dor stars and suggests that these pulsators do not reveal strong residual surface GIW flux variations compared to those of the prewhitened modes or waves.
For the SPB stars, however, \citet{VanBeeck2021} found 4 of the 26 SPB stars to have light curves with outbursts, similar to those observed in pulsating Be stars and interpreted in terms of GIWs \citep[e.g.,][]{Neiner2012,Neiner2020}. Following \citet{VanBeeck2021} we divide the 26 SPB pulsators into 3 groups: 
\begin{enumerate}
    \item 
13 SPB stars for which the significant coherent modes with constant amplitude and frequency gave a good fit to the measured light curve; these have $A_{\rm res}/A_{\rm coh} \in  [0.97,13.79]\%$;
\item 
4 SPB stars with Be-like outbursts due to modes with time-variable behaviour, in addition to coherent modes with constant amplitude in their light curve; these have $A_{\rm res}/A_{\rm coh} \in [18.92,59.04]\%$;
\item
9 SPB stars with unresolved large-amplitude coherent modes left in the residuals; for those we cannot derive a proper estimate for any residual GIW flux variability at the surface.
\end{enumerate}
Although we have only 17 SPB stars for which we can deduce a meaningful estimate of possible residual stochastic GIWs at the surface, a trend occurs. The residual amplitude of the 13 non-outbursting SPB stars with resolved modes is limited to less than 14\% of the dominant resonant-mode amplitude; these 13 stars cover rotation rates between 4\% and 88\% of the critical rate. The four outbursting SPB stars rotate between 72\% and 96\% of their critical rate and their residual amplitude is a considerable fraction (up to 59\%) of the significant prehwitened modes/waves. This suggests that outbursting SPB stars have variability that is partly due to time-dependent amplitudes and/or frequencies and may point to the co-existence of resonant modes driven in their envelope and GIW triggered locally by the rotating convective core and surviving radiative damping while travelling to the surface, as suggested for outbursting Be pulsators \citep{Neiner2020}.

To get a first physical understanding of these observational results, we revisited the theoretical work by \cite{Augustson2020}. They studied the stochastic excitation of GIWs by the Reynolds stresses in turbulent rotating convective regions as a function of the convective Rossby number and of the stiffness of the convective/radiative interface. Using their model, we  computed the GIW flux ($F_z$) normalised by the gravity-wave flux in the non-rotating case ($F_0$) for values of the stiffness typical of SPB stars (i.e.\ $S=10^3$) and of $\gamma\,$Dor stars (i.e. $S=10^4$). The results are shown in 
the left and right panels of Fig.\,\ref{fig:wavefluxAMA20}, respectively.
For convective Rossby numbers observed for SPB stars (${\rm Ro}_{\rm c}$ between $10^{-2.3}$ and $10^{-0.8}$), we see that the ratio of the wave flux can vary between $10^{-5}$ and $10^5$. This opens the path to observe stochastically-excited GIWs at the surface of these stars if those waves can propagate through the near-surface layers. For convective Rossby numbers observed for $\gamma\,$Dor stars (${\rm Ro}_{\rm c}$ between $10^{-3}$ and $10^{-1.5}$), we find that the ratios of the wave fluxes are often below unity, making a detection for such stars more difficult, except for stars that rotate slowly such that ${\rm Ro}_{\rm c}\gtrsim 10^{-2}$. About one third of our sample of $\gamma\,$Dor stars fall between this lower bound and ${\rm Ro}_{\rm c}\sim 10^{-1.5}$.  Thus, according to the local Cartesian analytical model with a monochromatic wave excitation by small-scale turbulent convective Reynolds stresses derived in \cite{Augustson2020}, subinertial GIWs could have sufficient energy flux to be observed with an amplitude between 1 and 100 times the one expected for pure gravity waves. These trends are in good agreement with observations \citep{Bowman2019a,Bowman2019b,Bowman2020}. To confirm these results, however, devoted theoretical predictions and numerical simulations in spherical geometry should be computed, building on relevant work such as \citet{Edelmann2019} and \citet{Neiner2020} and using the calibrated stiffness parameters and Rossby numbers as calibrated in this work.

\section{Differential rotation and magnetic field in the convective core}

\begin{figure}
\begin{center} 
\rotatebox{270}{\resizebox{6.4cm}{!}{\includegraphics{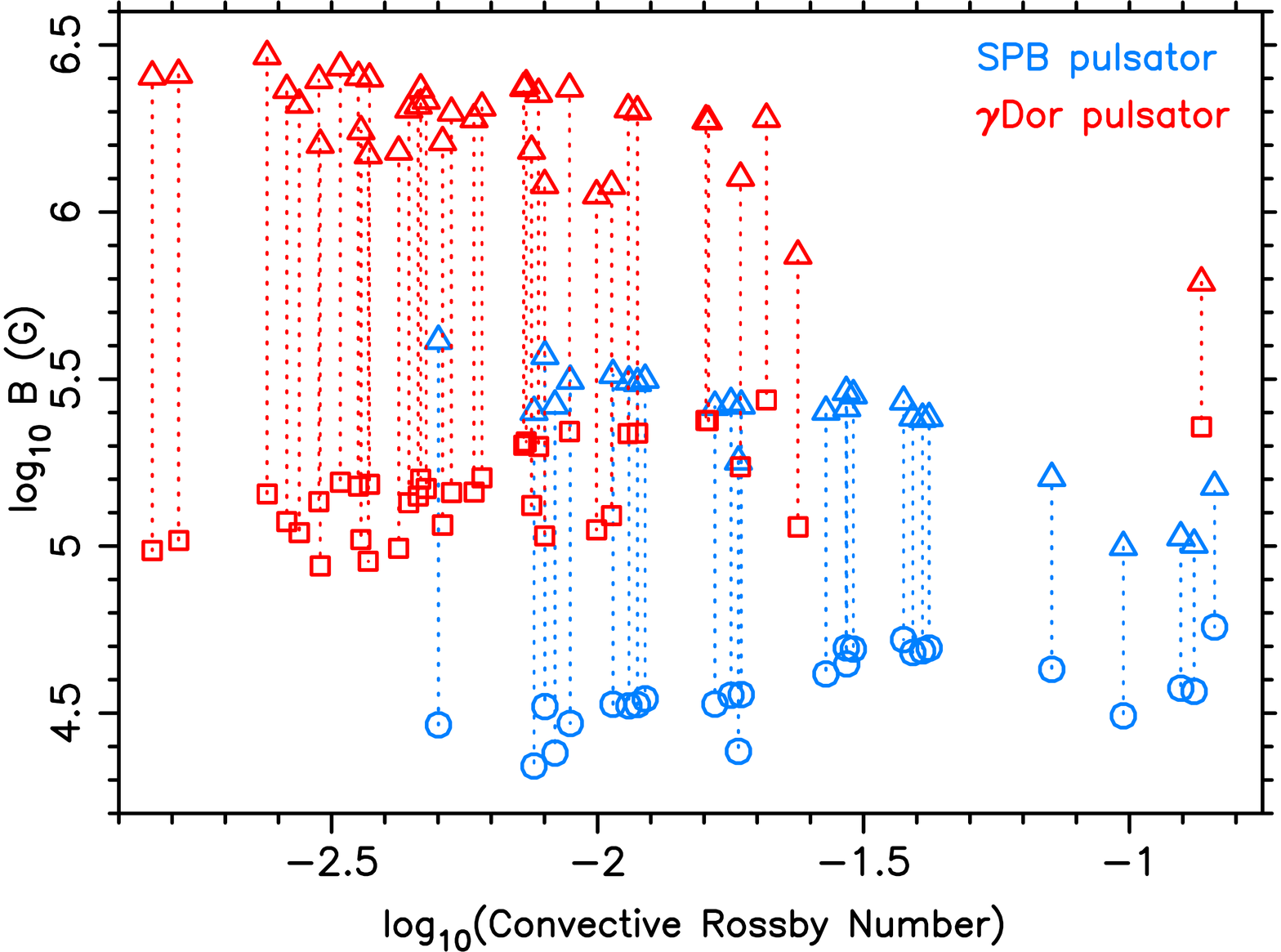}}}
\caption{Estimates of the magnetic field following equipartition ($B\propto \sqrt{\rm KE}$) and magnetostrophic ($B\propto \sqrt{{\rm KE}/{\rm Ro}_{\rm c}}$) scalings as in \cite{Augustson2019b}, with ${\rm KE}$ being the mixing length value of the kinetic energy at the radius where $\nu_{\mathrm{conv}}$ is taken from each stellar structure model.  The red squares/triangles denote the equipartition/magnetostropic
bounds of the estimates for the $\gamma\,$Dor stars and the blue circles/triangles those for the SPB stars, with the two bounding values connected via a dotted line for each star in the two samples.}\label{fig:magfield}
\end{center}
\end{figure}

Rotational properties of convective cores can in principle be probed thanks to subinertial gravito-inertial modes, which become propagative inertial modes
\citep{Ouazzani2020,Saio2021}. Following this recently discovered potential, 
it is interesting to discuss the expected differential rotation and magnetic states of convective cores. 
For stars with convective Rossby numbers below unity, turbulent (magneto-)convection should be strongly constrained by rotation. In such a  configuration, the differential rotation is expected to be mostly cylindrical because of the so-called Taylor-Proudman constraint \citep[cf.\,][]{Garaud2009}. In addition, the degree of differential rotation should decrease with increasing strength of the magnetic field \citep{Brun2005,Augustson2016}.

Adopting such a framework, we provide estimates of the strength of the magnetic field triggered by the dynamo action due to the turbulent interaction of rotating convection and the magnetic field. On the one hand, in the regime where the convective Rossby number is below unity, we expect the field to be generated in the magnetostrophic regime, where the Lorentz force balances the Coriolis acceleration as formalised by \citet{Augustson2019b}. This is seen in numerical simulations such as those by \citet{Featherstone2009} and \citet{Augustson2016}, as well as in the simulations of rotating convective dynamos described in \citet{Augustson2019b}, \citet{calkins21}, and \citet{orvedahl21}.  On the other hand, the dynamos could be in a subequipartition or equipartition regime, where the magnetic and kinetic energies are proportional to each other, that is a state like the one in which the Sun appears to be \citep{brownbp11,augustson15,brun17}. 

Applying dynamo scaling laws as derived by \cite{Augustson2019b}, that is $B_{\rm equi}^2=8\pi\rho\,v_{\rm conv}^2$ and 
$B_{\rm stroph}^2=8\pi\rho\,v_{\rm conv}^2/{\rm Ro}_{\rm c}$,
we find peak magnetic field intensities driven by convective dynamo action in the stellar cores ranging from $20$ to $400$\,kG in the SPB stars and from $0.1$ to $3$~MG in the $\gamma\,$Dor stars, as illustrated in Fig.\,\ref{fig:magfield}.
Such core magnetic fields are likely hidden from the surface for two reasons:
\begin{enumerate}
    \item 
Rapid drop off of the magnetic field is expected to be induced by small-scale currents, which decay approximately as $r_{\rm cc}^{\ell}/r^{\ell+2}$, where $r_{\rm cc}$ is the convective core radius, $r$ the local radius outside the core, and $\ell$ the angular degree of the sectoral mode approximating the morphology of the convection. This result is obtained via a multipolar expansion of the integral form of the Maxwell equations \citep[e.g.,][]{jefimenko92}.  
\item
Even if loops of the field can escape, they will take a significant fraction of the lifetime of the star to reach the surface \citep[e.g.,][]{macgregor03}.
\end{enumerate}
However, pulsations propagating in the core are expected to be impacted by such a field and may offer the opportunity to get detected via magnetically-induced frequency splittings \citep{Prat2019,Mathis2021}, which is the subject of an ongoing study (Augustson \& Mathis, in prep.).

\section{Conclusions}

Our work provides stringent asteroseismic calibrations of key quantities that
are usually included as free parameters in 3D numerical simulations of stellar
interiors. The asteroseismic calibrations are delivered by detected and
identified { low-frequency prograde} gravito-inertial modes deduced from {\it
  Kepler\/} space photometry. These modes are direct and unique probes of
the core/envelope interface region in rotating dwarfs. Previous studies of
asteroseismic calibrations for Rossby numbers so far focused on pressure modes
detected in slowly-rotating low-mass sun-like stars. The latest study by
\citet{Corsaro2021} revealed envelope convective turnover time scales of
low-mass dwarfs between $\sim\!10$ and $\sim\!55$ days. These estimates resulted
from taking averages over the entire slowly-rotating convective envelope and
using asteroseismic scalings by mass, radius, and luminosity while also relying
on colours of the stars.  This led to convective Rossby numbers roughly between
0.05 and 0.5 (we note that we use a different definition of Rossby numbers by a
factor two in this paper, as we rely on the expression and definition derived
from the momentum equation).  Our study focused on a completely different regime
of the parameter space. Moreover, we provide a direct localised estimate of
Rossby numbers from identified { prograde} 
modes, which probe the physical properties
in the transition region between the convective core and the radiative envelope
of rapidly rotating stars. Our sample, covering masses from 1.3 to 9\,M$_\odot$
and rotation rates from almost zero to the critical rate, leads to convective
Rossby numbers roughly between 0.002 for the fastest rotators and 0.15 for the
slowest rotators.

Asteroseismic modelling so far revealed that $\gamma\,$Dor stars have  lower levels of envelope mixing than SPB stars \citep[][Table\,1, for a summary]{Aerts2021}. While this result needs further scrutiny by including both shear mixing \citep[as has been done for the SPB stars, see\ ][]{Pedersen2021} and radiative levitation \citep{Mombarg2020} in the envelope of $\gamma\,$Dor stars, this result is consistent with our current findings.
Indeed, the inferred Rossby numbers derived from asteroseismic modelling reveals that the local GIW flux at the convective core boundary expected for SPB stars can reach higher values than the ones anticipated for $\gamma\,$Dor stars \citep[see Fig.\,\ref{fig:wavefluxAMA20} and also][Fig.\,6]{Augustson2020}. If, aside from angular momentum transport, GIWs are also (partly) responsible for the transport of chemical elements in the stellar envelope as derived by \citet{Rogers2017}, it is more efficient for SPB stars than for $\gamma\,$Dor stars. This is exactly what asteroseismic inferences from observations show. In addition, we have demonstrated that the convective penetration length as a function of the convective Rossby number as predicted in \cite{Augustson2019} is in good agreement with the results found from the best forward asteroseismic models for the 17 SPB stars in our sample for which \citet{Pedersen2021} found penetration to be the better explanation than diffusive exponentially decaying  core overshooting. 

% The sample size of our study was limited to only 63 stars. Due to this
% limitation, and also due to the lack of a single homogeneous forward modelling
% approach for the sample as a whole, we focused our work on three pertinent
% time scales, that is $\nu_{\rm rot}$, $N_{\rm max}$, and
% $\nu_{\rm char}^{\rm conv}$. With considerably larger samples, it would become
% meaningful to perform multivariate studies to assess the correlation between
% the stiffness values, Rossby numbers, penetration sizes, and core magnetic
% field predictions and other global stellar parameters, such as the stellar
% mass, luminosity, convective core mass, metallicity, evolutionary stage,
% angular momentum, etc. We anticipate such studies to become possible from a
% much larger sample covering the entire upper main sequence from long-term
% ($>$\,3 years) TESS extended mission \citep{Jenkins2020} and/or PLATO
% \citep{Rauer2014,Montalto2021} light curves of gravity-mode pulsators.

\cite{Augustson2019} provided the first coherent theoretical description of convective penetration while taking into account the Coriolis acceleration. This formalism meanwhile provided a good explanation for the mixing of light elements such as Lithium in low-mass stars \citep{Dumont2021b,Dumont2021a}. Here, we conclude in addition that this theory of rotating convection along with its accompanied predictions for convective penetration \citep{Augustson2019} and GIW stochastic excitation \citep{Augustson2020} is also in agreement with asteroseismically calibrated Rossby numbers, stiffness values, and levels of envelope mixing inferred from gravity-mode pulsators \citep{Mombarg2021,Pedersen2021}. Our results clearly
demonstrate the need to take into account the Coriolis acceleration in stellar modelling, while it has so far mostly been ignored in state-of-the-art structure and evolution models of rotating stars. We therefore suggest to include these theories and their asteroseismically calibrated parameter values as provided here in new generations of stellar evolution models \citep[e.g.][]{Rieutord2013,Gagnier2019}.

Finally, thanks to the convective Rossby numbers calibrated from GIW asteroseismology as done in this work, we open new windows to probe the rotational and magnetic properties of rotating convective cores. Indeed, it is highly probable that rapid rotators have cylindrical differential rotation whose contrast will depend on the strength of dynamo-generated magnetic fields. We predict amplitudes for such fields ranging from $20$ to $400$~kG for SPB stars and from $0.1$ to $3$~MG for $\gamma\,$Dor stars. Such high-amplitude magnetic fields should be detectable via asteroseismology \citep{Prat2020,VanBeeck2020,Mathis2021,Bugnet2021}.

\begin{acknowledgements}
  The research leading to these results has received funding from the European
  Research Council (ERC) under the European Union’s Horizon 2020 research and
  innovation programme (grant agreements N$^\circ$670519: MAMSIE with PI Aerts
  and N$^\circ$647383: SPIRE with PI Mathis) and from the KU\,Leuven Research
  Council (grant C16/18/005: PARADISE). CA acknowledges support from the BELgian
  federal Science Policy Office (BELSPO) through a PRODEX grant for the ESA
  space mission PLATO, while SM acknowledges support from the CNES PLATO grant
  at CEA/DAP. This research was supported in part by the National Science
  Foundation under Grant No. NSF PHY-1748958. { TVR and JSGM gratefully
    acknowledge support from the Research Foundation Flanders (FWO) 
under grant
    agreements 12ZB620N and V429020N, respectively.}  The authors are thankful
  to Tami Rogers and her group for interesting and stimulating discussions
  during biannual informal workshops among the Leuven, Saclay, and Newcastle
  teams the past years.  CA also acknowledges useful discussions with Raphael
  Hirschi, Dave Arnett, Daniel Lecoanet, and Matteo Cantiello, as well as
  helpful comments on the draft paper prior to submission from Dominic Bowman
  and Cole Johnston. { We appreciate the positive comments
    received from the referee and the suggestion to use a more appropriate
    terminology for the gravito-inertial mode identification.}
\end{acknowledgements}

\bibliographystyle{aa} % style aa.bst
\bibliography{GIW.bib} % your references Yourfile.bib

\begin{thebibliography}{110}
\expandafter\ifx\csname natexlab\endcsname\relax\def\natexlab#1{#1}\fi

\bibitem[{{Aerts}(2021)}]{Aerts2021}
{Aerts}, C. 2021, Reviews of Modern Physics, 93, 015001

\bibitem[{{Aerts} {et~al.}(2010){Aerts}, {Christensen-Dalsgaard}, \&
  {Kurtz}}]{Aerts2010}
{Aerts}, C., {Christensen-Dalsgaard}, J., \& {Kurtz}, D.~W. 2010,
  {Asteroseismology, Springer-Verlag Heidelberg}

\bibitem[{{Aerts} {et~al.}(2019){Aerts}, {Mathis}, \& {Rogers}}]{Aerts2019}
{Aerts}, C., {Mathis}, S., \& {Rogers}, T.~M. 2019, \araa, 57, 35

\bibitem[{{Aerts} {et~al.}(2018){Aerts}, {Molenberghs}, {Michielsen},
  {Pedersen}, {Bj{\"o}rklund}, {Johnston}, {Mombarg}, {Bowman}, {Buysschaert},
  {P{\'a}pics}, {Sekaran}, {Sundqvist}, {Tkachenko}, {Truyaert}, {Van Reeth},
  \& {Vermeyen}}]{Aerts2018}
{Aerts}, C., {Molenberghs}, G., {Michielsen}, M., {et~al.} 2018, \apjs, 237, 15

\bibitem[{{Aerts} {et~al.}(2017){Aerts}, {Van Reeth}, \&
  {Tkachenko}}]{Aerts2017}
{Aerts}, C., {Van Reeth}, T., \& {Tkachenko}, A. 2017, \apjl, 847, L7

\bibitem[{{Alvan} {et~al.}(2014){Alvan}, {Brun}, \& {Mathis}}]{Alvan2014}
{Alvan}, L., {Brun}, A.~S., \& {Mathis}, S. 2014, \aap, 565, A42

\bibitem[{{Alvan} {et~al.}(2015){Alvan}, {Strugarek}, {Brun}, {Mathis}, \&
  {Garcia}}]{Alvan2015}
{Alvan}, L., {Strugarek}, A., {Brun}, A.~S., {Mathis}, S., \& {Garcia}, R.~A.
  2015, \aap, 581, A112

\bibitem[{{Andr\'e}(2019)}]{Andre2019}
{Andr\'e}, Q. 2019, PhD thesis, Universit\'e Paris-Saclay, Universit\'e Paris
  Diderot, Sorbonne Paris Cit\'e, France

\bibitem[{{Augustson} {et~al.}(2015){Augustson}, {Brun}, {Miesch}, \&
  {Toomre}}]{augustson15}
{Augustson}, K., {Brun}, A.~S., {Miesch}, M., \& {Toomre}, J. 2015, \apj, 809,
  149

\bibitem[{{Augustson} {et~al.}(2016){Augustson}, {Brun}, \&
  {Toomre}}]{Augustson2016}
{Augustson}, K.~C., {Brun}, A.~S., \& {Toomre}, J. 2016, \apj, 829, 92

\bibitem[{{Augustson} {et~al.}(2019){Augustson}, {Brun}, \&
  {Toomre}}]{Augustson2019b}
{Augustson}, K.~C., {Brun}, A.~S., \& {Toomre}, J. 2019, \apj, 876, 83

\bibitem[{{Augustson} \& {Mathis}(2019)}]{Augustson2019}
{Augustson}, K.~C. \& {Mathis}, S. 2019, \apj, 874, 83

\bibitem[{{Augustson} {et~al.}(2020){Augustson}, {Mathis}, \&
  {Astoul}}]{Augustson2020}
{Augustson}, K.~C., {Mathis}, S., \& {Astoul}, A. 2020, \apj, 903, 90

\bibitem[{{Auvergne} {et~al.}(2009){Auvergne}, {Bodin}, {Boisnard}, {Buey},
  {Chaintreuil}, {Epstein}, {Jouret}, {Lam-Trong}, {Levacher}, {Magnan},
  {Perez}, {Plasson}, {Plesseria}, {Peter}, {Steller}, {Tiph{\`e}ne}, {Baglin},
  {Agogu{\'e}}, {Appourchaux}, {Barbet}, {Beaufort}, {Bellenger}, {Berlin},
  {Bernardi}, {Blouin}, {Boumier}, {Bonneau}, {Briet}, {Butler}, {Cautain},
  {Chiavassa}, {Costes}, {Cuvilho}, {Cunha-Parro}, {de Oliveira Fialho},
  {Decaudin}, {Defise}, {Djalal}, {Docclo}, {Drummond}, {Dupuis}, {Exil},
  {Faur{\'e}}, {Gaboriaud}, {Gamet}, {Gavalda}, {Grolleau}, {Gueguen},
  {Guivarc'h}, {Guterman}, {Hasiba}, {Huntzinger}, {Hustaix}, {Imbert},
  {Jeanville}, {Johlander}, {Jorda}, {Journoud}, {Karioty}, {Kerjean},
  {Lafond}, {Lapeyrere}, {Landiech}, {Larqu{\'e}}, {Laudet}, {Le Merrer},
  {Leporati}, {Leruyet}, {Levieuge}, {Llebaria}, {Martin}, {Mazy}, {Mesnager},
  {Michel}, {Moalic}, {Monjoin}, {Naudet}, {Neukirchner}, {Nguyen-Kim},
  {Ollivier}, {Orcesi}, {Ottacher}, {Oulali}, {Parisot}, {Perruchot},
  {Piacentino}, {Pinheiro da Silva}, {Platzer}, {Pontet}, {Pradines},
  {Quentin}, {Rohbeck}, {Rolland}, {Rollenhagen}, {Romagnan}, {Russ}, {Samadi},
  {Schmidt}, {Schwartz}, {Sebbag}, {Smit}, {Sunter}, {Tello}, {Toulouse},
  {Ulmer}, {Vandermarcq}, {Vergnault}, {Wallner}, {Waultier}, \&
  {Zanatta}}]{Auvergne2009}
{Auvergne}, M., {Bodin}, P., {Boisnard}, L., {et~al.} 2009, \aap, 506, 411

\bibitem[{{B{\"o}hm-Vitense}(1958)}]{BohmVitense1958}
{B{\"o}hm-Vitense}, E. 1958, \zap, 46, 108

\bibitem[{{Bouabid} {et~al.}(2013){Bouabid}, {Dupret}, {Salmon},
  {Montalb{\'a}n}, {Miglio}, \& {Noels}}]{Bouabid2013}
{Bouabid}, M.~P., {Dupret}, M.~A., {Salmon}, S., {et~al.} 2013, \mnras, 429,
  2500

\bibitem[{{Bowman} {et~al.}(2019{\natexlab{a}}){Bowman}, {Aerts}, {Johnston},
  {Pedersen}, {Rogers}, {Edelmann}, {Sim{\'o}n-D{\'\i}az}, {Van Reeth},
  {Buysschaert}, {Tkachenko}, \& {Triana}}]{Bowman2019a}
{Bowman}, D.~M., {Aerts}, C., {Johnston}, C., {et~al.} 2019{\natexlab{a}},
  \aap, 621, A135

\bibitem[{{Bowman} {et~al.}(2019{\natexlab{b}}){Bowman}, {Burssens},
  {Pedersen}, {Johnston}, {Aerts}, {Buysschaert}, {Michielsen}, {Tkachenko},
  {Rogers}, {Edelmann}, {Ratnasingam}, {Sim{\'o}n-D{\'\i}az}, {Castro},
  {Moravveji}, {Pope}, {White}, \& {De Cat}}]{Bowman2019b}
{Bowman}, D.~M., {Burssens}, S., {Pedersen}, M.~G., {et~al.}
  2019{\natexlab{b}}, Nature Astronomy, 3, 760

\bibitem[{{Bowman} {et~al.}(2020){Bowman}, {Burssens}, {Sim{\'o}n-D{\'\i}az},
  {Edelmann}, {Rogers}, {Horst}, {R{\"o}pke}, \& {Aerts}}]{Bowman2020}
{Bowman}, D.~M., {Burssens}, S., {Sim{\'o}n-D{\'\i}az}, S., {et~al.} 2020,
  \aap, 640, A36

\bibitem[{{Brown} {et~al.}(2011){Brown}, {Miesch}, {Browning}, {Brun}, \&
  {Toomre}}]{brownbp11}
{Brown}, B.~P., {Miesch}, M.~S., {Browning}, M.~K., {Brun}, A.~S., \& {Toomre},
  J. 2011, \apj, 731, 69

\bibitem[{{Browning} {et~al.}(2004){Browning}, {Brun}, \&
  {Toomre}}]{Browning2004}
{Browning}, M.~K., {Brun}, A.~S., \& {Toomre}, J. 2004, \apj, 601, 512

\bibitem[{{Brun} \& {Browning}(2017)}]{brun17}
{Brun}, A.~S. \& {Browning}, M.~K. 2017, Living Reviews in Solar Physics, 14, 4

\bibitem[{{Brun} {et~al.}(2005){Brun}, {Browning}, \& {Toomre}}]{Brun2005}
{Brun}, A.~S., {Browning}, M.~K., \& {Toomre}, J. 2005, \apj, 629, 461

\bibitem[{{Buchler} {et~al.}(1997){Buchler}, {Goupil}, \&
  {Hansen}}]{Buchler1997}
{Buchler}, J.~R., {Goupil}, M.~J., \& {Hansen}, C.~J. 1997, \aap, 321, 159

\bibitem[{{Bugnet} {et~al.}(2021){Bugnet}, {Prat}, {Mathis}, {Astoul},
  {Augustson}, {Garc{\'\i}a}, {Mathur}, {Amard}, \& {Neiner}}]{Bugnet2021}
{Bugnet}, L., {Prat}, V., {Mathis}, S., {et~al.} 2021, \aap, 650, A53

\bibitem[{{Buysschaert} {et~al.}(2018){Buysschaert}, {Aerts}, {Bowman},
  {Johnston}, {Van Reeth}, {Pedersen}, {Mathis}, \& {Neiner}}]{Buysschaert2018}
{Buysschaert}, B., {Aerts}, C., {Bowman}, D.~M., {et~al.} 2018, \aap, 616, A148

\bibitem[{{Calkins} {et~al.}(2021){Calkins}, {Orvedahl}, \&
  {Featherstone}}]{calkins21}
{Calkins}, M.~A., {Orvedahl}, R.~J., \& {Featherstone}, N.~A. 2021, Geophysical
  Journal International

\bibitem[{{Choi} {et~al.}(2016){Choi}, {Dotter}, {Conroy}, {Cantiello},
  {Paxton}, \& {Johnson}}]{Choi2016}
{Choi}, J., {Dotter}, A., {Conroy}, C., {et~al.} 2016, \apj, 823, 102

\bibitem[{{Christophe} {et~al.}(2018){Christophe}, {Ballot}, {Ouazzani},
  {Antoci}, \& {Salmon}}]{Christophe2018}
{Christophe}, S., {Ballot}, J., {Ouazzani}, R.~M., {Antoci}, V., \& {Salmon},
  S.~J.~A.~J. 2018, \aap, 618, A47

\bibitem[{{Claeskens} \& {Hjort}(2008)}]{Claeskens2008}
{Claeskens}, G. \& {Hjort}, N.~L. 2008, {Model Selection and Model Averaging,
  Cambridge Series in Statistical and Probabilistic Mathematics}

\bibitem[{{Corsaro} {et~al.}(2021){Corsaro}, {Bonanno}, {Mathur},
  {Garc{\'\i}a}, {Santos}, {Breton}, \& {Khalatyan}}]{Corsaro2021}
{Corsaro}, E., {Bonanno}, A., {Mathur}, S., {et~al.} 2021, \aap, 652, L2

\bibitem[{{Cristini} {et~al.}(2019){Cristini}, {Hirschi}, {Meakin}, {Arnett},
  {Georgy}, \& {Walkington}}]{Christini2019}
{Cristini}, A., {Hirschi}, R., {Meakin}, C., {et~al.} 2019, \mnras, 484, 4645

\bibitem[{{Degroote} {et~al.}(2010){Degroote}, {Aerts}, {Baglin}, {Miglio},
  {Briquet}, {Noels}, {Niemczura}, {Montalban}, {Bloemen}, {Oreiro},
  {Vu{\v{c}}kovi{\'c}}, {Smolders}, {Auvergne}, {Baudin}, {Catala}, \&
  {Michel}}]{Degroote2010}
{Degroote}, P., {Aerts}, C., {Baglin}, A., {et~al.} 2010, \nat, 464, 259

\bibitem[{{Dhouib} {et~al.}(2021{\natexlab{a}}){Dhouib}, {Prat}, {Van Reeth},
  \& {Mathis}}]{Dhouib2021a}
{Dhouib}, H., {Prat}, V., {Van Reeth}, T., \& {Mathis}, S. 2021{\natexlab{a}},
  \aap, 652, A154

\bibitem[{{Dhouib} {et~al.}(2021{\natexlab{b}}){Dhouib}, {Prat}, {Van Reeth},
  \& {Mathis}}]{Dhouib2021b}
{Dhouib}, H., {Prat}, V., {Van Reeth}, T., \& {Mathis}, S. 2021{\natexlab{b}},
  \aap, in press, arXiv:2110.03619

\bibitem[{{Dintrans} \& {Rieutord}(2000)}]{Dintrans2000}
{Dintrans}, B. \& {Rieutord}, M. 2000, \aap, 354, 86

\bibitem[{{Dumont} {et~al.}(2021{\natexlab{a}}){Dumont}, {Charbonnel},
  {Palacios}, \& {Borisov}}]{Dumont2021b}
{Dumont}, T., {Charbonnel}, C., {Palacios}, A., \& {Borisov}, S.
  2021{\natexlab{a}}, \aap, 654, A46

\bibitem[{{Dumont} {et~al.}(2021{\natexlab{b}}){Dumont}, {Palacios},
  {Charbonnel}, {Richard}, {Amard}, {Augustson}, \& {Mathis}}]{Dumont2021a}
{Dumont}, T., {Palacios}, A., {Charbonnel}, C., {et~al.} 2021{\natexlab{b}},
  \aap, 646, A48

\bibitem[{{Dupret} {et~al.}(2005){Dupret}, {Grigahc{\`e}ne}, {Garrido},
  {Gabriel}, \& {Scuflaire}}]{Dupret2005}
{Dupret}, M.~A., {Grigahc{\`e}ne}, A., {Garrido}, R., {Gabriel}, M., \&
  {Scuflaire}, R. 2005, \aap, 435, 927

\bibitem[{{Edelmann} {et~al.}(2019){Edelmann}, {Ratnasingam}, {Pedersen},
  {Bowman}, {Prat}, \& {Rogers}}]{Edelmann2019}
{Edelmann}, P.~V.~F., {Ratnasingam}, R.~P., {Pedersen}, M.~G., {et~al.} 2019,
  \apj, 876, 4

\bibitem[{{Featherstone} {et~al.}(2009){Featherstone}, {Browning}, {Brun}, \&
  {Toomre}}]{Featherstone2009}
{Featherstone}, N.~A., {Browning}, M.~K., {Brun}, A.~S., \& {Toomre}, J. 2009,
  \apj, 705, 1000

\bibitem[{{Fuller}(2017)}]{Fuller2017}
{Fuller}, J. 2017, \mnras, 472, 1538

\bibitem[{{Gagnier} {et~al.}(2019){Gagnier}, {Rieutord}, {Charbonnel},
  {Putigny}, \& {Espinosa Lara}}]{Gagnier2019}
{Gagnier}, D., {Rieutord}, M., {Charbonnel}, C., {Putigny}, B., \& {Espinosa
  Lara}, F. 2019, \aap, 625, A89

\bibitem[{{Garaud} \& {Acevedo Arreguin}(2009)}]{Garaud2009}
{Garaud}, P. \& {Acevedo Arreguin}, L. 2009, \apj, 704, 1

\bibitem[{{Gebruers} {et~al.}(2021){Gebruers}, {Straumit}, {Tkachenko},
  {Mombarg}, {Pedersen}, {Van Reeth}, {Li}, {Lampens}, {Escorza}, {Bowman}, {De
  Cat}, {Vermeylen}, {Bodensteiner}, {Rix}, \& {Aerts}}]{Gebruers2021}
{Gebruers}, S., {Straumit}, I., {Tkachenko}, A., {et~al.} 2021, \aap, 650, A151

\bibitem[{{Georgy} {et~al.}(2013){Georgy}, {Ekstr{\"o}m}, {Granada}, {Meynet},
  {Mowlavi}, {Eggenberger}, \& {Maeder}}]{Georgy2013}
{Georgy}, C., {Ekstr{\"o}m}, S., {Granada}, A., {et~al.} 2013, \aap, 553, A24

\bibitem[{{Goupil} \& {Buchler}(1994)}]{Goupil1994}
{Goupil}, M.-J. \& {Buchler}, J.~R. 1994, \aap, 291, 481

\bibitem[{{Guzik} {et~al.}(2000){Guzik}, {Kaye}, {Bradley}, {Cox}, \&
  {Neuforge}}]{Guzik2000}
{Guzik}, J.~A., {Kaye}, A.~B., {Bradley}, P.~A., {Cox}, A.~N., \& {Neuforge},
  C. 2000, \apjl, 542, L57

\bibitem[{{Henneco} {et~al.}(2021){Henneco}, {Van Reeth}, {Prat}, {Mathis},
  {Mombarg}, \& {Aerts}}]{Henneco2021}
{Henneco}, J., {Van Reeth}, T., {Prat}, V., {et~al.} 2021, \aap, 648, A97

\bibitem[{{Horst} {et~al.}(2020){Horst}, {Edelmann}, {Andr{\'a}ssy},
  {R{\"o}pke}, {Bowman}, {Aerts}, \& {Ratnasingam}}]{Horst2020}
{Horst}, L., {Edelmann}, P.~V.~F., {Andr{\'a}ssy}, R., {et~al.} 2020, \aap,
  641, A18

\bibitem[{{Jefimenko}(1992)}]{jefimenko92}
{Jefimenko}, O.~D. 1992, American Journal of Physics, 60, 899

\bibitem[{{Johnston} {et~al.}(2019){Johnston}, {Tkachenko}, {Aerts},
  {Molenberghs}, {Bowman}, {Pedersen}, {Buysschaert}, \&
  {P{\'a}pics}}]{Johnston2019}
{Johnston}, C., {Tkachenko}, A., {Aerts}, C., {et~al.} 2019, \mnras, 482, 1231

\bibitem[{{Koch} {et~al.}(2010){Koch}, {Borucki}, {Basri}, {Batalha}, {Brown},
  {Caldwell}, {Christensen-Dalsgaard}, {Cochran}, {DeVore}, {Dunham},
  {Gautier}, {Geary}, {Gilliland}, {Gould}, {Jenkins}, {Kondo}, {Latham},
  {Lissauer}, {Marcy}, {Monet}, {Sasselov}, {Boss}, {Brownlee}, {Caldwell},
  {Dupree}, {Howell}, {Kjeldsen}, {Meibom}, {Morrison}, {Owen}, {Reitsema},
  {Tarter}, {Bryson}, {Dotson}, {Gazis}, {Haas}, {Kolodziejczak}, {Rowe}, {Van
  Cleve}, {Allen}, {Chandrasekaran}, {Clarke}, {Li}, {Quintana}, {Tenenbaum},
  {Twicken}, \& {Wu}}]{Koch2010}
{Koch}, D.~G., {Borucki}, W.~J., {Basri}, G., {et~al.} 2010, \apjl, 713, L79

\bibitem[{{Kurtz} {et~al.}(2014){Kurtz}, {Saio}, {Takata}, {Shibahashi},
  {Murphy}, \& {Sekii}}]{Kurtz2014}
{Kurtz}, D.~W., {Saio}, H., {Takata}, M., {et~al.} 2014, \mnras, 444, 102

\bibitem[{{Lecoanet} {et~al.}(2019){Lecoanet}, {Cantiello}, {Quataert},
  {Couston}, {Burns}, {Pope}, {Jermyn}, {Favier}, \& {Le Bars}}]{Lecoanet2019}
{Lecoanet}, D., {Cantiello}, M., {Quataert}, E., {et~al.} 2019, \apjl, 886, L15

\bibitem[{{Lee}(2021)}]{Lee2021}
{Lee}, U. 2021, \mnras, 505, 1495

\bibitem[{{Lee} \& {Saio}(1987)}]{LeeSaio1987}
{Lee}, U. \& {Saio}, H. 1987, \mnras, 224, 513

\bibitem[{{Lee} \& {Saio}(1997)}]{LeeSaio1997}
{Lee}, U. \& {Saio}, H. 1997, \apj, 491, 839

\bibitem[{{Lee} \& {Saio}(2020)}]{Lee2020}
{Lee}, U. \& {Saio}, H. 2020, \mnras, 497, 4117

\bibitem[{{Li} {et~al.}(2019){Li}, {Bedding}, {Murphy}, {Van Reeth}, {Antoci},
  \& {Ouazzani}}]{GangLi2019}
{Li}, G., {Bedding}, T.~R., {Murphy}, S.~J., {et~al.} 2019, \mnras, 482, 1757

\bibitem[{{Li} {et~al.}(2020){Li}, {Van Reeth}, {Bedding}, {Murphy}, {Antoci},
  {Ouazzani}, \& {Barbara}}]{GangLi2020}
{Li}, G., {Van Reeth}, T., {Bedding}, T.~R., {et~al.} 2020, \mnras, 491, 3586

\bibitem[{{MacGregor} \& {Cassinelli}(2003)}]{macgregor03}
{MacGregor}, K.~B. \& {Cassinelli}, J.~P. 2003, \apj, 586, 480

\bibitem[{{Mathis}(2009)}]{Mathis2009}
{Mathis}, S. 2009, \aap, 506, 811

\bibitem[{{Mathis} {et~al.}(2021){Mathis}, {Bugnet}, {Prat}, {Augustson},
  {Mathur}, \& {Garcia}}]{Mathis2021}
{Mathis}, S., {Bugnet}, L., {Prat}, V., {et~al.} 2021, \aap, 647, A122

\bibitem[{{Mathis} {et~al.}(2014){Mathis}, {Neiner}, \& {Tran
  Minh}}]{Mathis2014}
{Mathis}, S., {Neiner}, C., \& {Tran Minh}, N. 2014, \aap, 565, A47

\bibitem[{{Mathis} \& {Prat}(2019)}]{MathisPrat2019}
{Mathis}, S. \& {Prat}, V. 2019, \aap, 631, A26

\bibitem[{{Michielsen} {et~al.}(2019){Michielsen}, {Pedersen}, {Augustson},
  {Mathis}, \& {Aerts}}]{Michielsen2019}
{Michielsen}, M., {Pedersen}, M.~G., {Augustson}, K.~C., {Mathis}, S., \&
  {Aerts}, C. 2019, \aap, 628, A76

\bibitem[{{Miglio} {et~al.}(2008){Miglio}, {Montalb{\'a}n}, {Noels}, \&
  {Eggenberger}}]{Miglio2008}
{Miglio}, A., {Montalb{\'a}n}, J., {Noels}, A., \& {Eggenberger}, P. 2008,
  \mnras, 386, 1487

\bibitem[{{Mombarg} {et~al.}(2020){Mombarg}, {Dotter}, {Van Reeth},
  {Tkachenko}, {Gebruers}, \& {Aerts}}]{Mombarg2020}
{Mombarg}, J. S.~G., {Dotter}, A., {Van Reeth}, T., {et~al.} 2020, \apj, 895,
  51

\bibitem[{{Mombarg} {et~al.}(2021){Mombarg}, {Van Reeth}, \&
  {Aerts}}]{Mombarg2021}
{Mombarg}, J.~S.~G., {Van Reeth}, T., \& {Aerts}, C. 2021, \aap, 650, A58

\bibitem[{{Mombarg} {et~al.}(2019){Mombarg}, {Van Reeth}, {Pedersen},
  {Molenberghs}, {Bowman}, {Johnston}, {Tkachenko}, \& {Aerts}}]{Mombarg2019}
{Mombarg}, J.~S.~G., {Van Reeth}, T., {Pedersen}, M.~G., {et~al.} 2019, \mnras,
  485, 3248

\bibitem[{{Moravveji} {et~al.}(2016){Moravveji}, {Townsend}, {Aerts}, \&
  {Mathis}}]{Moravveji2016}
{Moravveji}, E., {Townsend}, R. H.~D., {Aerts}, C., \& {Mathis}, S. 2016, \apj,
  823, 130

\bibitem[{{Neiner} {et~al.}(2012){Neiner}, {Floquet}, {Samadi}, {Espinosa
  Lara}, {Fr{\'e}mat}, {Mathis}, {Leroy}, {de Batz}, {Rainer}, {Poretti},
  {Mathias}, {Guarro Fl{\'o}}, {Buil}, {Ribeiro}, {Alecian}, {Andrade},
  {Briquet}, {Diago}, {Emilio}, {Fabregat}, {Guti{\'e}rrez-Soto}, {Hubert},
  {Janot-Pacheco}, {Martayan}, {Semaan}, {Suso}, \& {Zorec}}]{Neiner2012}
{Neiner}, C., {Floquet}, M., {Samadi}, R., {et~al.} 2012, \aap, 546, A47

\bibitem[{{Neiner} {et~al.}(2020){Neiner}, {Lee}, {Mathis}, {Saio}, {Lovekin},
  \& {Augustson}}]{Neiner2020}
{Neiner}, C., {Lee}, U., {Mathis}, S., {et~al.} 2020, \aap, 644, A9

\bibitem[{{Orvedahl} {et~al.}(2021){Orvedahl}, {Featherstone}, \&
  {Calkins}}]{orvedahl21}
{Orvedahl}, R.~J., {Featherstone}, N.~A., \& {Calkins}, M.~A. 2021, \mnras,
  507, L67

\bibitem[{{Ouazzani} {et~al.}(2020){Ouazzani}, {Ligni{\`e}res}, {Dupret},
  {Salmon}, {Ballot}, {Christophe}, \& {Takata}}]{Ouazzani2020}
{Ouazzani}, R.~M., {Ligni{\`e}res}, F., {Dupret}, M.~A., {et~al.} 2020, \aap,
  640, A49

\bibitem[{{Ouazzani} {et~al.}(2019){Ouazzani}, {Marques}, {Goupil},
  {Christophe}, {Antoci}, {Salmon}, \& {Ballot}}]{Ouazzani2019}
{Ouazzani}, R.~M., {Marques}, J.~P., {Goupil}, M.~J., {et~al.} 2019, \aap, 626,
  A121

\bibitem[{{Ouazzani} {et~al.}(2017){Ouazzani}, {Salmon}, {Antoci}, {Bedding},
  {Murphy}, \& {Roxburgh}}]{Ouazzani2017}
{Ouazzani}, R.-M., {Salmon}, S.~J.~A.~J., {Antoci}, V., {et~al.} 2017, \mnras,
  465, 2294

\bibitem[{{Pamyatnykh}(1999)}]{Pamyatnykh1999}
{Pamyatnykh}, A.~A. 1999, \actaa, 49, 119

\bibitem[{{P{\'a}pics} {et~al.}(2017){P{\'a}pics}, {Tkachenko}, {Van Reeth},
  {Aerts}, {Moravveji}, {Van de Sande}, {De Smedt}, {Bloemen}, {Southworth},
  {Debosscher}, {Niemczura}, \& {Gameiro}}]{Papics2017}
{P{\'a}pics}, P.~I., {Tkachenko}, A., {Van Reeth}, T., {et~al.} 2017, \aap,
  598, A74

\bibitem[{{Pedersen}(2021)}]{Pedersen2021b}
{Pedersen}, M.~G. 2021, \apj, submitted

\bibitem[{{Pedersen} {et~al.}(2021){Pedersen}, {Aerts}, {P{\'a}pics},
  {Michielsen}, {Gebruers}, {Rogers}, {Molenberghs}, {Burssens}, {Garcia}, \&
  {Bowman}}]{Pedersen2021}
{Pedersen}, M.~G., {Aerts}, C., {P{\'a}pics}, P.~I., {et~al.} 2021, Nature
  Astronomy, 5, 715

\bibitem[{{Pedersen} {et~al.}(2018){Pedersen}, {Aerts}, {P{\'a}pics}, \&
  {Rogers}}]{Pedersen2018}
{Pedersen}, M.~G., {Aerts}, C., {P{\'a}pics}, P.~I., \& {Rogers}, T.~M. 2018,
  \aap, 614, A128

\bibitem[{{Prat} {et~al.}(2019){Prat}, {Mathis}, {Buysschaert}, {Van Beeck},
  {Bowman}, {Aerts}, \& {Neiner}}]{Prat2019}
{Prat}, V., {Mathis}, S., {Buysschaert}, B., {et~al.} 2019, \aap, 627, A64

\bibitem[{{Prat} {et~al.}(2020){Prat}, {Mathis}, {Neiner}, {Van Beeck},
  {Bowman}, \& {Aerts}}]{Prat2020}
{Prat}, V., {Mathis}, S., {Neiner}, C., {et~al.} 2020, \aap, 636, A100

\bibitem[{{Rieutord} \& {Espinosa Lara}(2013)}]{Rieutord2013}
{Rieutord}, M. \& {Espinosa Lara}, F. 2013, in EAS Publications Series,
  Vol.~63, EAS Publications Series, ed. G.~{Alecian}, Y.~{Lebreton},
  O.~{Richard}, \& G.~{Vauclair}, 385--394

\bibitem[{{Rogers} {et~al.}(2013){Rogers}, {Lin}, {McElwaine}, \&
  {Lau}}]{Rogers2013}
{Rogers}, T.~M., {Lin}, D.~N.~C., {McElwaine}, J.~N., \& {Lau}, H.~H.~B. 2013,
  \apj, 772, 21

\bibitem[{{Rogers} \& {McElwaine}(2017)}]{Rogers2017}
{Rogers}, T.~M. \& {McElwaine}, J.~N. 2017, \apjl, 848, L1

\bibitem[{{Saio} {et~al.}(2018){Saio}, {Kurtz}, {Murphy}, {Antoci}, \&
  {Lee}}]{Saio2018}
{Saio}, H., {Kurtz}, D.~W., {Murphy}, S.~J., {Antoci}, V.~L., \& {Lee}, U.
  2018, \mnras, 474, 2774

\bibitem[{{Saio} {et~al.}(2015){Saio}, {Kurtz}, {Takata}, {Shibahashi},
  {Murphy}, {Sekii}, \& {Bedding}}]{Saio2015}
{Saio}, H., {Kurtz}, D.~W., {Takata}, M., {et~al.} 2015, \mnras, 447, 3264

\bibitem[{{Saio} {et~al.}(2021){Saio}, {Takata}, {Lee}, {Li}, \& {Van
  Reeth}}]{Saio2021}
{Saio}, H., {Takata}, M., {Lee}, U., {Li}, G., \& {Van Reeth}, T. 2021, \mnras,
  502, 5856

\bibitem[{{Samadi} {et~al.}(2010){Samadi}, {Belkacem}, {Goupil}, {Dupret},
  {Brun}, \& {Noels}}]{Samadi2010}
{Samadi}, R., {Belkacem}, K., {Goupil}, M.~J., {et~al.} 2010, \apss, 328, 253

\bibitem[{{Schmid} \& {Aerts}(2016)}]{Schmid2016}
{Schmid}, V.~S. \& {Aerts}, C. 2016, \aap, 592, A116

\bibitem[{{Scott} {et~al.}(2021){Scott}, {Hirschi}, {Georgy}, {Arnett},
  {Meakin}, {Kaiser}, {Ekstr{\"o}m}, \& {Yusof}}]{Scott2021}
{Scott}, L.~J.~A., {Hirschi}, R., {Georgy}, C., {et~al.} 2021, \mnras, 503,
  4208

\bibitem[{{Szewczuk} \& {Daszy{\'n}ska-Daszkiewicz}(2017)}]{Szewczuk2017}
{Szewczuk}, W. \& {Daszy{\'n}ska-Daszkiewicz}, J. 2017, \mnras, 469, 13

\bibitem[{{Szewczuk} \& {Daszy{\'n}ska-Daszkiewicz}(2018)}]{Szewczuk2018}
{Szewczuk}, W. \& {Daszy{\'n}ska-Daszkiewicz}, J. 2018, \mnras, 478, 2243

\bibitem[{{Szewczuk} {et~al.}(2021){Szewczuk}, {Walczak}, \&
  {Daszy{\'n}ska-Daszkiewicz}}]{Szewczuk2021}
{Szewczuk}, W., {Walczak}, P., \& {Daszy{\'n}ska-Daszkiewicz}, J. 2021, \mnras,
  503, 5894

\bibitem[{{Takata} {et~al.}(2020){Takata}, {Ouazzani}, {Saio}, {Christophe},
  {Ballot}, {Antoci}, {Salmon}, \& {Hijikawa}}]{Takata2020}
{Takata}, M., {Ouazzani}, R.~M., {Saio}, H., {et~al.} 2020, \aap, 635, A106

\bibitem[{{Tassoul}(1980)}]{Tassoul1980}
{Tassoul}, M. 1980, \apjs, 43, 469

\bibitem[{{Tkachenko} {et~al.}(2013){Tkachenko}, {Aerts}, {Yakushechkin},
  {Debosscher}, {Degroote}, {Bloemen}, {P{\'a}pics}, {de Vries}, {Lombaert},
  {Hrudkova}, {Fr{\'e}mat}, {Raskin}, \& {Van Winckel}}]{Tkachenko2013}
{Tkachenko}, A., {Aerts}, C., {Yakushechkin}, A., {et~al.} 2013, \aap, 556, A52

\bibitem[{{Townsend}(2003)}]{Townsend2003}
{Townsend}, R.~H.~D. 2003, \mnras, 340, 1020

\bibitem[{{Townsend}(2020)}]{Townsend2020}
{Townsend}, R.~H.~D. 2020, \mnras, 497, 2670

\bibitem[{{Triana} {et~al.}(2015){Triana}, {Moravveji}, {P{\'a}pics}, {Aerts},
  {Kawaler}, \& {Christensen-Dalsgaard}}]{Triana2015}
{Triana}, S.~A., {Moravveji}, E., {P{\'a}pics}, P.~I., {et~al.} 2015, \apj,
  810, 16

\bibitem[{{Unno} {et~al.}(1989){Unno}, {Osaki}, {Ando}, {Saio}, \&
  {Shibahashi}}]{Unno1989}
{Unno}, W., {Osaki}, Y., {Ando}, H., {Saio}, H., \& {Shibahashi}, H. 1989,
  {Nonradial oscillations of stars, Tokyo: University of Tokyo Press, 1989, 2nd
  ed.}

\bibitem[{{Van Beeck} {et~al.}(2021){Van Beeck}, {Bowman}, {Pedersen}, {Van
  Reeth}, {Van Hoolst}, \& {Aerts}}]{VanBeeck2021}
{Van Beeck}, J., {Bowman}, D.~M., {Pedersen}, M.~G., {et~al.} 2021, \aap, in
  press, arXiv:2108.02907

\bibitem[{{Van Beeck} {et~al.}(2020){Van Beeck}, {Prat}, {Van Reeth}, {Mathis},
  {Bowman}, {Neiner}, \& {Aerts}}]{VanBeeck2020}
{Van Beeck}, J., {Prat}, V., {Van Reeth}, T., {et~al.} 2020, \aap, 638, A149

\bibitem[{{Van Reeth} {et~al.}(2016){Van Reeth}, {Tkachenko}, \&
  {Aerts}}]{VanReeth2016}
{Van Reeth}, T., {Tkachenko}, A., \& {Aerts}, C. 2016, \aap, 593, A120

\bibitem[{{Van Reeth} {et~al.}(2015){Van Reeth}, {Tkachenko}, {Aerts},
  {P{\'a}pics}, {Triana}, {Zwintz}, {Degroote}, {Debosscher}, {Bloemen},
  {Schmid}, {De Smedt}, {Fremat}, {Fuentes}, {Homan}, {Hrudkova},
  {Karjalainen}, {Lombaert}, {Nemeth}, {{\O}stensen}, {Van De Steene}, {Vos},
  {Raskin}, \& {Van Winckel}}]{VanReeth2015}
{Van Reeth}, T., {Tkachenko}, A., {Aerts}, C., {et~al.} 2015, \apjs, 218, 27

\bibitem[{{Wu} \& {Li}(2019)}]{Wu2019}
{Wu}, T. \& {Li}, Y. 2019, \apj, 881, 86

\bibitem[{{Wu} {et~al.}(2018){Wu}, {Li}, \& {Deng}}]{Wu2018}
{Wu}, T., {Li}, Y., \& {Deng}, Z.-m. 2018, \apj, 867, 47

\end{thebibliography}

\end{document}